\documentclass[12pt,preprint]{aastex}

\shorttitle{NLTE Lick indices for XMP stars}
\shortauthors{Short, Young \& Layden}

\begin{document}


\title{NLTE and LTE Lick indices for red giants from $[{{\rm M}\over {\rm H}}]$ 0.0 to -6.0 at SDSS and IDS spectral resolution}


\author{C. Ian Short}
\affil{Department of Astronomy \& Physics and Institute for Computational Astrophysics, Saint Mary's University,
    Halifax, NS, Canada, B3H 3C3}
\email{ishort@ap.smu.ca}

\author{Mitchell E. Young}
\affil{Department of Astronomy \& Physics and Institute for Computational Astrophysics, Saint Mary's University,
    Halifax, NS, Canada, B3H 3C3}
\email{}

\author{Nicholas Layden}
\affil{Department of Astronomy \& Physics and Institute for Computational Astrophysics, Saint Mary's University,
    Halifax, NS, Canada, B3H 3C3}
\email{}




\begin{abstract}

We investigate the dependence of the complete system of 22 Lick indices on overall metallicity scaled from solar
abundances, $[{{\rm M}\over {\rm H}}]$,
from the solar value, 0.0, down to the extremely-metal-poor (XMP) value of -6.0, for late-type giant stars 
(MK luminosity class III, $\log g=2.0$) of MK spectral class late-K to late-F ($3750 < T_{\rm eff} < 6500$ K) of the type
that are detected as ``fossils'' of early galaxy formation in the Galactic halo and in extra-galactic structures. 
Our investigation is based on synthetic index values, $I$, derived from atmospheric models and synthetic spectra
computed with PHOENIX in LTE and Non-LTE (NLTE), where the synthetic spectra have been convolved to the spectral resolution, $R$,
of both IDS and SDSS (and LAMOST) spectroscopy. 
We identify nine indices, that we designate ``Lick-XMP'', that remain both detectable and significantly 
$[{{\rm M}\over {\rm H}}]$-dependent down to $[{{\rm M}\over {\rm H}}]$ values of at least $\sim -5.0$,
and down to $[{{\rm M}\over {\rm H}}]$ $\sim -6.0$ in five cases, while also 
remaining well-behaved (single-valued as a function of $[{{\rm M}\over {\rm H}}]$ and positive in linear units). 
For these nine, we study the dependence of $I$ on NLTE effects, and on spectral resolution.  For our 
LTE $I$ values for spectra of SDSS resolution, we present the fitted polynomial coefficients, $C_{\rm n}$, from 
multi-variate linear regression for $I$ with terms up to third order in the independent variable pairs 
($T_{\rm eff}$, $[{{\rm M}\over {\rm H}}]$), 
and ($V-K$, $[{{\rm M}\over {\rm H}}]$), and compare them
to the fitted $C_{\rm n}$ values of \citet{worthey94} at IDS spectral resolution.  For this fitted $I$ data-set we 
present tables of  LTE partial derivatives,
${{\partial I}\over {\partial T_{\rm eff}}}|_{[{{\rm M}/{\rm H}}]}$,
${{\partial I}\over {\partial [{{\rm M}/{\rm H}}]}}|_{T{\rm eff}}$,
${{\partial I}\over {\partial (V-K)}}|_{[{{\rm M}/{\rm H}}]}$,
and ${{\partial I}\over {\partial [{{\rm M}/{\rm H}}]}}|_{(V-K)}$,
that can be used to infer the relation between a given difference, $\Delta I$, and a difference  
$\Delta T_{\rm eff}$ or $\Delta (V-K)$, or a difference $\Delta [{{\rm M}\over {\rm H}}]$,
while the other parameters are held fixed. 
For Fe-dominated Lick indices, the effect of NLTE is to generally weaken the value of $I$ 
at any give $T_{\rm eff}$ and $[{{\rm M}\over {\rm H}}]$ values.  
As an example of the impact on stellar parameter estimation, for late-type giants of inferred 
$T_{\rm eff} \gtrsim 4200$ K, an Fe-dominated $I$ value 
computed in LTE that is too strong might be compensated for by inferring a $T_{\rm eff}$ value that
is too large.

\end{abstract}


\keywords{stars: atmospheres, fundamental parameters, late-type }

\section{Introduction}

The determination of stellar parameters, especially overall metallicity (denoted here $[{{\rm M}\over {\rm H}}]$
unless otherwise indicated) and detailed
abundances of individual metals, in stars of remote
Galactic and extra-galactic structures has become crucial to the study of galaxy formation and evolution, including that
of the Milky Way.  Metallicity distribution functions for galaxies and globular star clusters (GCs) reveal information
about multiple populations from multiple star formation episodes and allow the investigation of the history of star formation.
Chemical tagging of stellar populations allows the investigation of the link between Galactic structures such as GCs and
nearby extra-galactic ones such as ultra-faint dwarf (UFD) and dwarf spheroidal (dSphe) satellite galaxies, and the process
through which galaxies are assembled in hierarchical structure formation (see \citet{mucciarelli13} for a recent example,
and \citet{frebel13} and \citet{belokurov13} for reviews).  
However, stars in remote 
structures are often only significantly and efficiently detectable with low- to moderate-resolution spectroscopy, such as that of the
Sloan Digital Sky Survey - Sloan Extension for Galactic Understanding and Exploration (SDSS-SEGUE), that precludes 
the measurement of individual spectral lines, and usefully accurate $[{{\rm M}\over {\rm H}}]$ values must be
obtained from observationally expensive follow-up spectroscopy at high spectral resolution ({\it e.g.} see \citet{aoki}).   
\citet{ivezic12} contains a recent authoritative review of how modest resolution spectroscopic surveys are 
revolutionizing our study of Galactic populations and leading to insights into Galactic formation.
As a result, there is active interest in novel methods for extracting $[{{\rm M}\over {\rm H}}]$ and $T_{\rm eff}$
values from low- to moderate-resolution data (see \citet{schlaufman} for a recent investigation of a method
based on IR molecular bands, and \citet{miller} for a very recent investigation of a 
photometric method based on SDSS $ugriz$ photometry.)

\paragraph{}

The original system of 11 \citep{gorgas93}, and then 21 (\citet{worthey94}, W94 henceforth), Lick/IDS spectral indices, $I$, 
was defined to optimize the determination
of {\bf $[{{\rm Fe}\over{\rm H}}]$} and, hence, age ($t$), from integrated light (IL) spectra of 
faint spatially unresolved {\it old} stellar populations ($-1 < [{{\rm M}\over{\rm H}}] < +0.5$, $0.5 < t < 20$ Gyr) 
dominated by G and K stars, obtained at low spectra resolution ($R < 1000$) in the
$\lambda 4000$ to $6200$ \AA~ spectral band.  
(\citet{wortheyo97} explored four new indices based on two definitions, each, of
H$\gamma$- and H$\delta$-centered features, but found them to be of limited usefulness - a result
consistent with our own investigation.) 
Because of the emphasis on old stellar populations,
such as that of  globular clusters (GCs) and ``early-type'' galaxies, red giant (RGB) stars are important 
contributors to the IL spectrum.  The indices were discovered by empirically identifying composite
spectral features in low $R$ spectra of Galactic and GC G and K stars that showed a significant 
and useful correlation with one of $[{{\rm M}\over{\rm H}}]$, $T_{\rm eff}$, or $\log g$ while
(hopefully) depending less sensitively on the other two.      
The original Lick/IDS system was defined with spectra obtained with the Lick Observatory Image Dissector Scanner (IDS) 
having a spectral sampling, $\Delta\lambda$, of 
$\sim 8$ \AA~ in a region centered at 5200 \AA~ spanning a $\Delta\lambda$ range of 2400 \AA, corresponding to 
spectral resolution $R \equiv \lambda / \Delta\lambda$ of $\sim 650$.  

\paragraph{}

Given the usefulness of the Lick indices for modern moderate resolution spectroscopic surveys such as SDSS-SEGUE and LAMOST, 
\citet{franchini10} developed the system further by creating a synthetic library of $I$ values for dwarf and giant stars 
derived from synthetic spectra
that had been convolved to the higher SDSS $R$ value of 1800, and tested the predicted relationship between the
$I$ values and stellar parameters against that of several empirical spectral libraries, including the SDSS-DR7 spectroscopic
database itself.  They also supplemented the 21 indices of W94 with a new near-UV index, namely CaHK - a prominent 
spectral feature that has proved useful for identifying candidate XMP stars in productive surveys such as
the HK survey of \citet{beers}.  
One of the main conclusions of \citet{franchini10} is that $T_{\rm eff}$ values derived from fitting their
synthetic indices to SDSS-SEGUE spectra of late-type giants were systematically lower than the $T_{\rm eff}$ 
values derived with the SEGUE Stellar Parameter Pipeline (SSPP).  


\paragraph{}

Many of the Lick indices are dominated by, or have significant contributions from, lines of \ion{Fe}{1}.  
Initial investigations of metal-poor stars in which the \ion{Fe}{1} extinction is treated with 
Non-local Thermodynamic Equilibrium (Non-LTE, NLTE) indicate that NLTE effects become increasingly important with decreasing metallicity, and alter the inferred $T_{\rm eff}$ 
values by as much as 400 K and the derived abundances of Fe by 0.4 dex for metal-poor giants with parameters based on RAVE 
survey spectra (\citet{serenelli13}, \citet{ruchti13}).  
Recently \citet{short13} and papers in that series have found that when most of the light and Fe-group metals that contribute to the
visible band spectral line blanketing of mildly metal-poor RGB stars is treated in NLTE, the $T_{\rm eff}$ value inferred from spectrophotometric spectral energy distributions (SEDs) of
$R \approx 100$ is reduced by $\gtrsim 50$ K.
Therefore, calibration of the Lick $I([{{\rm M}\over {\rm H}}])$
relation based on NLTE modeling may be crucial to using the indices for accurate $[{{\rm M}\over {\rm H}}]$ inference. 

\paragraph{}

 Our goal is to extend an analysis of the detectability, and sensitivity to stellar parameters, including 
$[{{\rm M}\over {\rm H}}]$, of the Lick indices to the regime of extremely metal-poor (XMP) red giants,
and to investigate the magnitude of NLTE effects on the value of modeled Lick indices.
In Section \ref{sModel} we describe the atmospheric models and the spectrum synthesis, and the procedure
for producing synthetic $I$ values; in Section \ref{sResults} we identify the Lick indices that remain most useful 
at XMP metallicities, and provide useful polynomial fits and partial derivatives for index values, $I$, modeled in LTE,
in terms of $T_{\rm eff}$, $V-K$, and $[{{\rm M}\over {\rm H}}]$; in Section \ref{sConc} we present 
conclusions.

\section{Modeling \label{sModel}}

\subsection{Model grid \label{sgrid}}

  We have used PHOENIX V. 15 \citep{hauschildtafba99} to computed a grid of atmospheric models and corresponding synthetic 
spectra in both LTE and ``massive multi-species NLTE'' \citep{shorth09} for very- and extremely-metal-poor (VMP 
and XMP) red (and ``orange'') giant stars
of MK spectral class late F to late K covering the range of $T_{\rm eff}$ and $[{{\rm M}\over {\rm H}}]$ value of stars that are 
spectroscopically accessible at Galactic halo distances and that serve as useful stellar ``fossils'' for Galactic archeology 
\citep{cohen13}.  
The grid also includes red giants of higher $[{{\rm M}\over {\rm H}}]$ value representative of the solar neighborhood
and disk population for comparison.  The parameters of the LTE grid are: \\

$3750 \leq T_{\rm eff} \leq 6500$ K with $\Delta T_{\rm eff} = 250$ K,  \\
$-6.0 \leq [{{\rm M}\over {\rm H}}] \leq 0.0$ with $\Delta [{{\rm M}\over {\rm H}}] = 1.0$ for $[{{\rm M}\over {\rm H}}] < -2.0$ and 0.5 for $[{{\rm M}\over {\rm H}}] \geq -2.0$, \\
$\log g = 2.0$\\

\paragraph{}  

\begin{figure}
\plotone{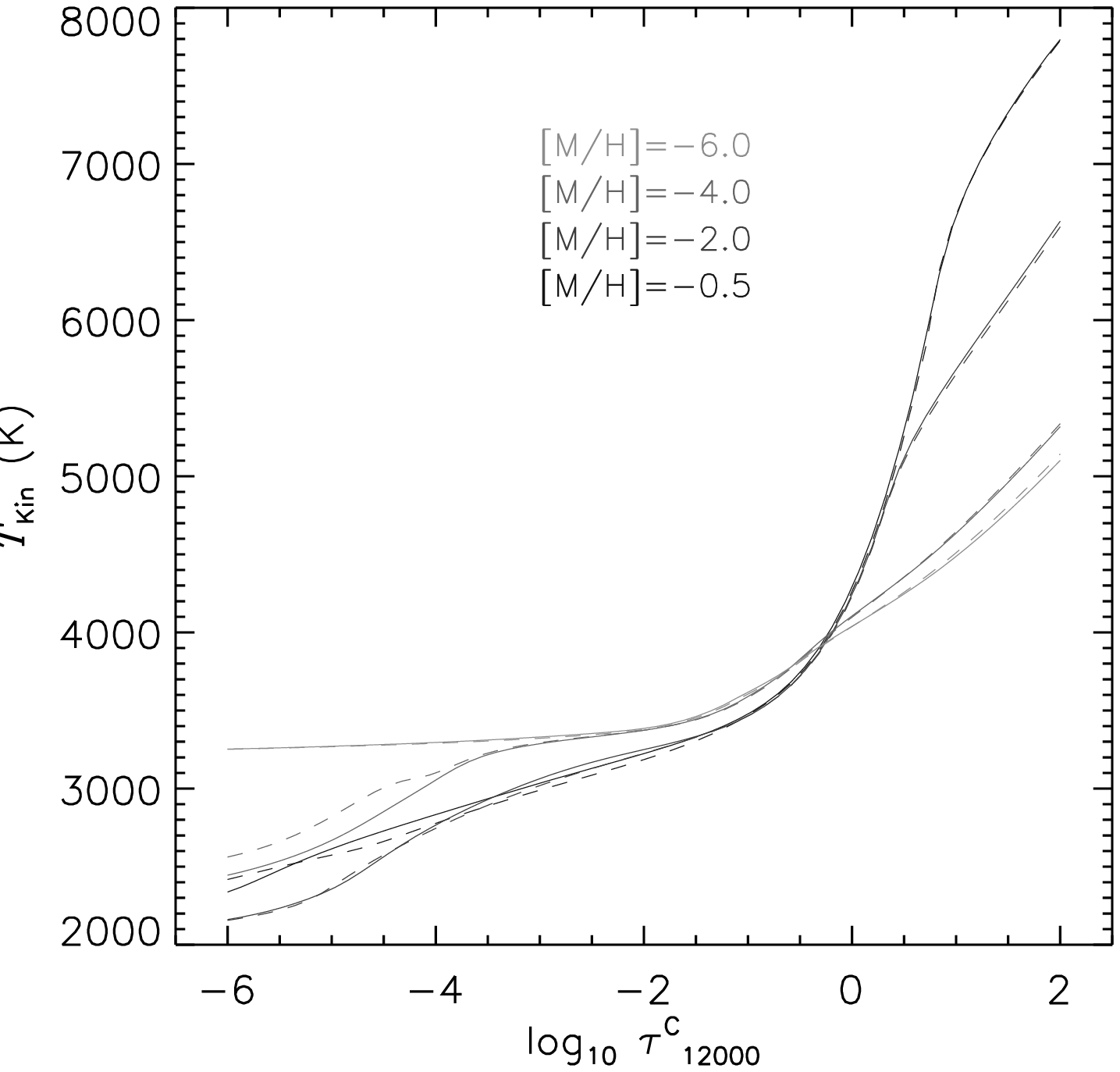}
\caption{$T_{\rm Kin}(\tau_{\rm 12000})$ structures for models of
$T_{\rm eff}=4000$ K and $[{{\rm M}\over{\rm H}}]$ values of -0.5, -2.0,
-4.0, and -6.0.  Gray-scale: darker line indicates larger 
$[{{\rm M}\over{\rm H}}]$ value. Results are shown for LTE (dashed line) 
and NLTE (solid line) models.  
  \label{atmosGrid}}
\end{figure}

Fig. \ref{atmosGrid} shows the $T_{\rm Kin}(\tau_{\rm 12000})$ structure of a subset of our models for $T_{\rm eff}=4000$ K 
and $[{{\rm M}\over {\rm H}}]$ values of -0.5, -2.0, -4.0, and -6.0, where $\tau_{\rm 12000}$ is the monochromatic
{\it continuum} optical depth at 12000 \AA~ and serves as our standard radial depth variable.  The reduction in the 
well-understood back-warming and surface cooling effects caused by line extinction as $[{{\rm M}\over {\rm H}}]$
decreases is readily noticeable.   
For comparison, the grid of \citet{franchini10} has $3500 \leq T_{\rm eff} \leq 7000$ K with 
$\Delta T_{\rm eff} = 250$ K, $-2.5 \leq [{{\rm Fe}\over {\rm H}}] \leq 0.5$ with $\Delta [{{\rm Fe}\over {\rm H}}] = 0.5$ generally,
with the addition of $[{{\rm Fe}\over {\rm H}}] = -4.0$ for their $\alpha$-enhanced models, and $0.5 \leq \log g \leq 5.0$ with 
$\Delta\log g = 0.5$, where $[{{\rm Fe}\over {\rm H}}]$ denotes the scaled abundance parameter for elements other than 
$\alpha$-process elements.  The most important distinguishing features of our grid are the extension to $[{{\rm M}\over {\rm H}}]$
values of -6.0, and the inclusion of NLTE models for a subset of parameter values spanning the grid.  
Although there are few stars of $[{{\rm M}\over {\rm H}}] \lesssim -3.5$, 
even among halo stars useful for Galactic archeology, extending the grid to $[{{\rm M}\over {\rm H}}] = -6.0$ allows us to
anchor the $I([{{\rm M}\over {\rm H}}])$ fit through the useful $[{{\rm M}\over {\rm H}}]$ range.  

\paragraph{}
 
  Because of the large number of $[{{\rm M}\over {\rm H}}]$ values (nine) and doubling of most of the grid to 
include NLTE counterparts, the number of models was limited by fixing the $\log g$ value at 2.0, representative of the giant population,
and giving all models a scaled-solar abundance distribution based on the abundances of \citet{grevs98} (therefore, for our grid
the overall metallicity parameter, $[{{\rm M}\over {\rm H}}]$, is identical with the $[{{\rm Fe}\over {\rm H}}]$ parameter).  W94
and \citet{gorgas93} found that important $T_{\rm eff}$- and {\bf $[{{\rm Fe}\over {\rm H}}]$-sensitive} $I$ values were 
least sensitive to $\log g$.  Nevertheless the lack of the $\log g$ dimension in our $I$ polynomial fits described
below leads to polynomial fitting coefficients that are not directly comparable to those of W94 from their
analysis of observed IDS spectra. 
Adopting an $\alpha$-enhancement of 0.0 is expected to have a minor effect on the differential comparison of most 
Fe-dominated $[{{\rm M}\over {\rm H}}]$-sensitive $I$ values
computed in LTE and NLTE (because these are really more $[{{\rm Fe}\over {\rm H}}]$ indicators, 
and Fe is not an $\alpha$-element), and will allow direct assessment of pure NLTE effects
on $I$ values across the entire range of $[{{\rm M}\over {\rm H}}]$ value.  Moreover, incorporating 
$\alpha$-enhancement requires a model for how the value of the enhancement increases with 
decreasing $[{{\rm M}\over {\rm H}}]$ value in the range 0.0 to -1.0.  Nevertheless, a future direction 
is to extend our grid in the $\alpha$-enhancement dimension.  As a preliminary assessment of the effect of
$\alpha$-enhancement, in Table \ref{alphaTab} we present the computed 
values of the index, $I$, computed at SDSS spectral resolution for our subset of nine ``Lick-XMP'' indices (see below) for LTE models
of $T_{\rm eff}=4000$ K, $\log g=2.0$, for which the $I$ values are relatively strong, and select 
$[{{\rm Fe}\over {\rm H}}]$ values of -2.0 and -4.0 and scaled
solar abundance, and abundances with the maximal relative enhancement of $+0.4$ of the eight $\alpha$-process elements of 
even atomic number from O ($Z=8$) to Ti ($Z=22$).  
Mg is an $\alpha$-process element, and the Mg$_{\rm 1}$ and  Mg $b$ indices
are more strongly affected, increasing by a factor of $\sim 1.5$ at $[{{\rm Fe}\over {\rm H}}]=-2$ 
where their value is still large.  Of our nine Lick-XMP indices, the results presented below for 
Mg$_{\rm 1}$ and  Mg $b$ should be regarded as most suspect and require follow up investigation 
with a full $\alpha$-enhanced NLTE and LTE model grids.  The inclusion of NLTE effects in
the modeling is not likely to change the conclusion that non-$\alpha$-element spectral features are
significantly less affected by $\alpha$-enhancement than are $\alpha$-element spectral features.   
 As might be expected, the effect of $\alpha$ enhancement on
the Fe-dominated indices Fe4531, Fe5015, {\bf Fe5270,} Fe5335, and the Na $D$ index is minor, being of
the order of $10\%$ or less.  {\bf We note that for some non-$\alpha$-element indices, the effect of 
$\alpha$-enhancement can be to {\it reduce} the index slightly at some $[{{\rm Fe}\over {\rm H}}]$-values, while   
increasing it at others.  Changes to the composition can affect the spectrum in the bracketing pseudo-continuum
windows that define any index, and will also indirectly affect the strength of features contributing to any index 
through the effect on the electron number density. } 

\paragraph{}

Our models have spherical geometry with radii based on an adopted mass of $M = 1 M_\odot$, a microturbulence broadening 
parameter of $\xi_{\rm T} = 2.0$ km s$^{\rm -1}$, which is consistent with what has been measured and adopted for late-type
giants generally, and a mixing-length parameter for the treatment of convective energy transport, $l$, of 
1.0 $H_{\rm P}$ (pressure scale heights). 

\subsubsection{NLTE treatment }

  We treat 6706 atomic energy-levels ($E$-levels) connected by 74550 bound-bound ($b-b$) transitions of 35 chemical 
species accounting for various ionization stages of 20 chemical elements, including H, He, CNO, and the 
Fe-group elements that blanket late-type visible band stellar spectra, as well as other abundant light metals. 
We compute the NLTE level populations, $n_{\rm i}(\tau_{\rm 12000})$, and hence the corresponding extinction coefficient, $\kappa_\lambda(\tau_{\rm 12000})$, in 
self-consistent multi-level
NLTE by solving the system of coupled rate equations of statistical equilibrium (SE) consistently with the equation of 
radiative transfer (RT) in each of the relevant bound-bound ($b-b$) and bound-free ($b-f$) transitions.  
\citet{shorth09} contains a description of
the species treated in NLTE, sources of atomic data, and other important details.

\paragraph{}

  Because NLTE models are more computationally expensive, we only produced NLTE models and spectra at a subset of 
our LTE grid, as follows\\ 

$4000 \leq T_{\rm eff} \leq 6500$ K with $\Delta T_{\rm eff} = 500$ K,  \\
$-6.0 \leq [{{\rm M}\over {\rm H}}] \leq 0.0$ with $\Delta [{{\rm M}\over {\rm H}}] = 1.0$, with addition of $[{{\rm M}\over {\rm H}}] = -0.5$ \\

This is sufficient to assess the dependence of NLTE effects on Lick indices throughout the grid.  Fig. \ref{atmosGrid} shows
NLTE $T_{\rm Kin}(\tau_{\rm 12000})$ structures for comparison with those of LTE.  NLTE radiative equilibrium is complex, and \citet{anderson89} contains
a very thorough analysis for the case of the Sun, and \citet{short12} 
extends the analysis to solar metallicity and moderately metal-poor RGB stars.    

\paragraph{NLTE Fe treatment }



  The predicted magnitude of the well-known \ion{Fe}{1} NLTE ``over-ionization'', and the resulting predicted brightening
of the \ion{Fe}{1}-blanketed near-UV and blue spectral bands with respect to the rest of the 
SED (see \citet{shorth09} and \cite{rutten86}) depends on
the details of the atomic model of \ion{Fe}{1} used in the NLTE Fe treatment.  More specifically, the completeness
with which high-energy $E$-levels are included near the ionization limit, $\chi_{\rm I}$, affects the computed rate of collisional recombination
from \ion{Fe}{2}, and thus the \ion{Fe}{1}/\ion{Fe}{2} ionization equilibrium \citep{mashonkina11}.  
Generally, the more $E$-levels are included for
which the atomic energy gap, $\Delta\chi$, between the $E$-level and $\chi_{\rm I}$ is less than the
average collisional energy among particles ($kT$) throughout the line-forming region of the atmosphere, the more accurate 
the NLTE effect on the computed SED will be.  
For \ion{Fe}{1}, $\chi_{\rm I} = 7.9024$ eV, and in our 494-level model \ion{Fe}{1} atom, 
the highest lying $E$-level has $\chi = 7.538$ eV, for a minimum $\Delta\chi$ gap of 0.364 eV.
The line-forming region of the atmosphere throughout the visible 
band generally lies at shallower total optical depths, $\tau_\lambda$, than the layer where the continuum value of $\tau_\lambda$ is unity,
where $T(\tau_\lambda) \le T_{\rm eff}$.  
For the warmest models in our grid 
($T_{\rm eff} = 6500$ K), $kT \le 0.560$ eV in the 
line forming region, and we have eight $E$-levels for which $\Delta\chi < kT$, at least in the lower
line-forming region. 
For the coolest models in our grid ($T_{\rm eff} =3750$ K), $kT \le 0.323$ eV in the 
line forming region, and we just miss having {\it any} $E$-levels for which $\Delta\chi < kT$.  
We expect our prediction of NLTE 
\ion{Fe}{1} effects to be most accurate at the warm end of our grid where collisional
recombination into our highest $E$-levels is energetically accessible. 
At the cool end, the collisional recombination rate is artificially suppressed by the lack
of higher-lying $E$-levels in the model atom.  The recombination rate is under-estimated for our cooler
models, and the NLTE over-ionization effect is 
likely over-estimated.  Our NLTE modeled NLTE effects on $I$ values may be thought of, cautiously,
as upper limits.  We plan to expand the PHOENIX NLTE \ion{Fe}{1}
atom in the near future, but this is a significant project in its own right. 

\subsection{Synthetic spectra \label{sspectra}}

  Our longer-term goal is to identify useful $T_{\rm eff}$ and overall $[{{\rm M}\over {\rm H}}]$ line diagnostics
for high spectral resolution from the near UV to the the near IR (NIR).  Therefore, we have computed 
synthetic spectra for each of our models for $3000 < \lambda < 26000$ \AA~ with a spectral sampling,
$\Delta\lambda$, set so as to maintain an $R$ value of ~300\, 000 throughout, sufficient to 
fully resolve spectral line cores.  This $\lambda$ range includes the NIR $J$, $H$, and $K$ photometric
bands, in which useful line lists of stellar parameter and abundance diagnostics have  recently been
published (\citet{bergemann12}, \citet{cesseti13}, \citet{le11}). 

\paragraph{}

\begin{figure}
\plotone{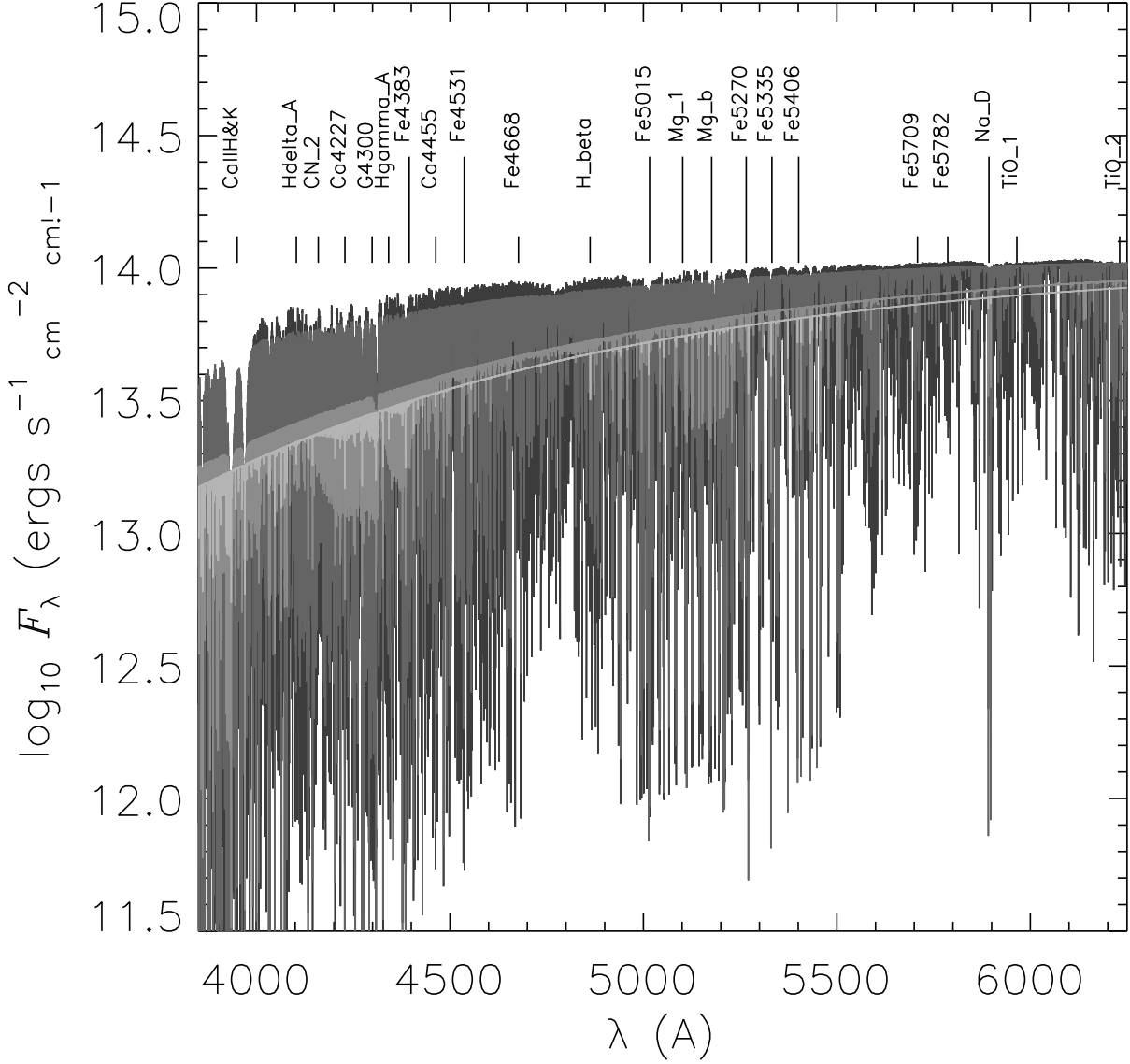}
\caption{Synthetic spectra for the LTE models of Fig. \ref{atmosGrid}, 
convolved to IDS spectral resolution for clarity.
The gray-scale is as in Fig. \ref{atmosGrid}.
Index labels that are offset upward refer to our nine Lick-XMP indices.
  \label{specGrid}}
\end{figure}

\begin{figure}
\plotone{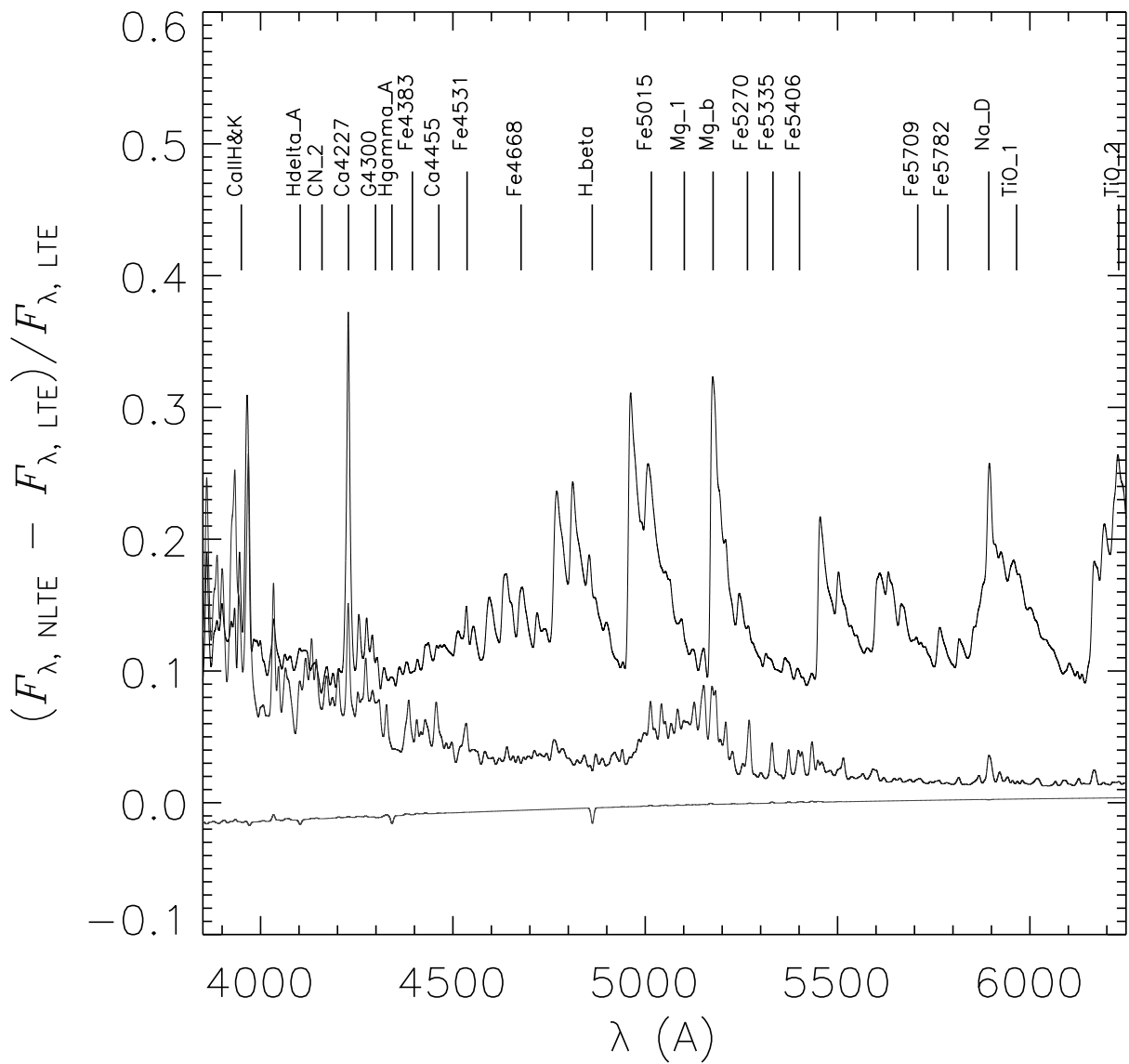}
\caption{Relative difference of synthetic spectra convolved to IDS resolution
 showing the comparison of NLTE to LTE results for models of $T_{\rm eff}=4000$ K and 
$[{{\rm M}\over{\rm H}}]$ values of -0.5, -2.0 and -6.0. 
The gray-scale and index label positions are as in Fig. \ref{specGrid}.
  \label{specDiff}}
\end{figure}

  Our synthetic spectra were post-processed by broadening with a Gaussian kernel to $R$ values of 
 650 and, following \citet{franchini10}, 1800 to match the resolution of the original IDS, and SDSS-SEGUE spectroscopy, 
respectively. 
A Gaussian is only an approximation to the real instrumental spectral profiles of IDS and SDSS-SEGUE
spectroscopy, but at this stage our study is a differential one to compare the effect of NLTE on Lick
indices to that of choice of $R$ value as a function of $[{{\rm M}\over {\rm H}}]$.
We do not account for either macro-turbulent or rotational broadening.  Rotation is expected to be modest in 
evolved stars of large radius, and both effects are expected to be minor at these $R$ values. 
Fig. \ref{specGrid} shows representative synthetic spectra for the models of  Fig. \ref{atmosGrid} convolved
to IDS spectral resolution, with the Lick indices labeled, and it can be seen 
that some of the strongest spectral features are still significant at 
$[{{\rm M}\over {\rm H}}]$ values as low as -6.0.  Fig. \ref{specDiff} shows the relative flux
difference at IDS resolution, 
$\Delta F_\lambda \equiv (F_{\lambda, {\rm NLTE}} - F_{\lambda, {\rm LTE}})/F_{\lambda_{\rm LTE}}$, for
models of $T_{\rm eff}= 4000$ K, and $[{{\rm M}\over {\rm H}}]$ values of 0.0, -2.0, and -6.0, with the Lick indices labeled.
$\Delta F_\lambda$ is generally positive at the $\lambda$ values of the Lick indices because most low-$\chi$
spectral lines from neutral ionization stages of metals are {\it weaker} in NLTE than in LTE.

\subsection{Synthetic Lick indices \label{sindex}}

  We use our LTE and NLTE synthetic spectra, convolved to both IDS and SDSS spectral resolution, to compute LTE and NLTE
IDS and SDSS Lick indices, $I$, following the prescription of W94.  We took the $\lambda$ values defining the latest 
recommended index and associated pseudo-continuum bands,
similar to the information presented in Table 1 of W94, from
the official Lick index WWW site (http://astro.wsu.edu/worthey/html/system.html), as did \citet{franchini10}.
The IDS indices conform to the 
well-studied Lick index system as originally defined, 
 and are directly comparable to 
those of \citet{gorgas93} and W94, and serve as a check on our procedures, as well as allowing us
to assess the impact of NLTE effects at IDS resolution.  The SDSS indices are comparable to those of \citet{franchini10}
and allow an assessment of NLTE effects at somewhat higher resolution typical of more modern spectroscopic
surveys such as SDSS and LAMOST.  Furthermore, comparing the LTE IDS and SDSS indices allows an assessment of the
dependence of the $I$ sensitivity to $T_{\rm eff}$ and $[{{\rm M}\over {\rm H}}]$ on $R$ value.   

\paragraph{}

  Following \citet{gorgas93} and W94, we also computed the photometric $V-K$ index for the models
as an observational surrogate for the independent parameter $T_{\rm eff}$.
  $V-K$ has been found to be {\it relatively} insensitive to $\log g$ and $[{{\rm M}\over{\rm H}}]$, and a good proxy for 
$T_{\rm eff}$ over the GK star range \citep{bellg89}.  
  For consistency with W94, we use the $V$- and $K$-band filter definitions of \citet{johnson66} and 
calibrate the index with a single-point calibration of the {\bf $[{{\rm Fe}\over {\rm H}}] = 0.0$} models at 
$T_{\rm eff} = 4000$ K to the $(V-K) - T_{\rm eff}$ 
relation given in Table 4 of \citet{ridgway80}.  The \citet{ridgway80} $(V-K) - T_{\rm eff}$ relation is 
for giant stars, and a $T_{\rm eff}$ value of 4000 K is near the center of their calibrated $T_{\rm eff}$ range,
and overlaps with the $T_{\rm eff}$ range of our grid.
We always use the LTE model $V-K$ color
on the grounds that it is serving as an independent variable in 
this analysis, and the LTE grid is more complete.

\paragraph{} 

  It is worth reiterating remarks made by previous investigators about the particular diagnostic utility of those indices 
that are expected to be most useful in this investigation:

\paragraph{} 

  Fe4383 and Fe4668 (W94) and Fe5270 and Fe5335 \citep{gorgas93} are $[{{\rm Fe}\over{\rm H}}]$-sensitive with a range in 
$I$ value significantly greater than measurement uncertainty, and are expected to be especially useful here if they remain detectable 
down to XMP metallicities (note that Fe4668 has significant contributions from Mg, Cr, Ti, and C$_{\rm 2}$).  Ca4227 really 
is dominated by Ca (whereas Ca4455 is more influenced by Fe-group lines), thus providing one of the few atomic indices 
{\it not} heavily affected by Fe, and is somewhat sensitive to overall $[{{\rm M}\over{\rm H}}]$ (W94) as well as 
$[{{\alpha}\over{\rm H}}]$ given that Ca is an $\alpha$-process element.  CN$_{\rm 2}$
is a modification of CN$_{\rm 1}$ designed to avoid contamination from the H$\delta$ line, and is strongly dependent on
overall $[{{\rm M}\over{\rm H}}]$ for giants (but less so for dwarfs) (W94).  
By contrast to the preceding, H$\beta$ has been found to be most
strongly dependent on $T_{\rm eff}$ \citep{gorgas93}, and thus provides a valuable complementary diagnostic.
Indices Mg$_{\rm 1}$ (dominated by MgH)
and Mg $b$ were found by \citet{gorgas93} to be usefully sensitive to $\log g$,
and we include them in our investigation to re-assess their $T_{\rm eff}$ and $[{{\rm M}\over{\rm H}}]$
sensitivity (note that Mg$_{\rm 2}$ includes contributions from both MgH 
and the Mg $b$ lines, so is less pure a signal of either).  Na $D$ is known to be significantly contaminated by interstellar (ISM) extinction,
which complicates its interpretation \citep{gorgas93}.   

\paragraph{} 

   \citet{franchini10} investigated the influence of enhanced $\alpha$-element abundances on model $I$ values. They 
found that for $T_{\rm eff} > 4250$ K the most $\alpha$-sensitive indices are CN$_{\rm 1}$, CN$_{\rm 2}$, 
CaHK, Ca4227, Fe4668, Mg$_{\rm 1}$, Mg$_{\rm 2}$, and Mg $b$, and that the least $\alpha$-sensitive indices are G4300, 
Ca4455, Fe4531, Fe5015, Fe5782, and H$\beta$.  For $T_{\rm eff} < 4250$ K the situation seems more complex, but CaHK, 
Ca4227 and Mg $b$ remain among the most $\alpha$-sensitive, and G4300 and Ca4455 remain among the least $\alpha$-sensitive
indices.

\section{Results \label{sResults}}

  We caution that because our models have scaled-solar abundances, in the discussion that follows the 
$[{{\rm M}\over{\rm H}}]$ parameter is effectively identical to the $[{{\rm Fe}\over{\rm H}}]$ parameter in the
internal context of our modeling and analysis, whereas
in $\alpha$-enhanced metal-poor RGB stars with non-solar abundance distributions, 
$[{{\rm M}\over{\rm H}}]$ differs from $[{{\rm Fe}\over{\rm H}}]$.  This distinction is expected to be most important
for those Lick indices that are Mg- or Ca- dominated, and less so for those that are
Fe-dominated. 

  \subsection{$V-K(T_{\rm eff})$ and $V-K([{{\rm M}\over{\rm H}}])$ relations }

\begin{figure}
\plotone{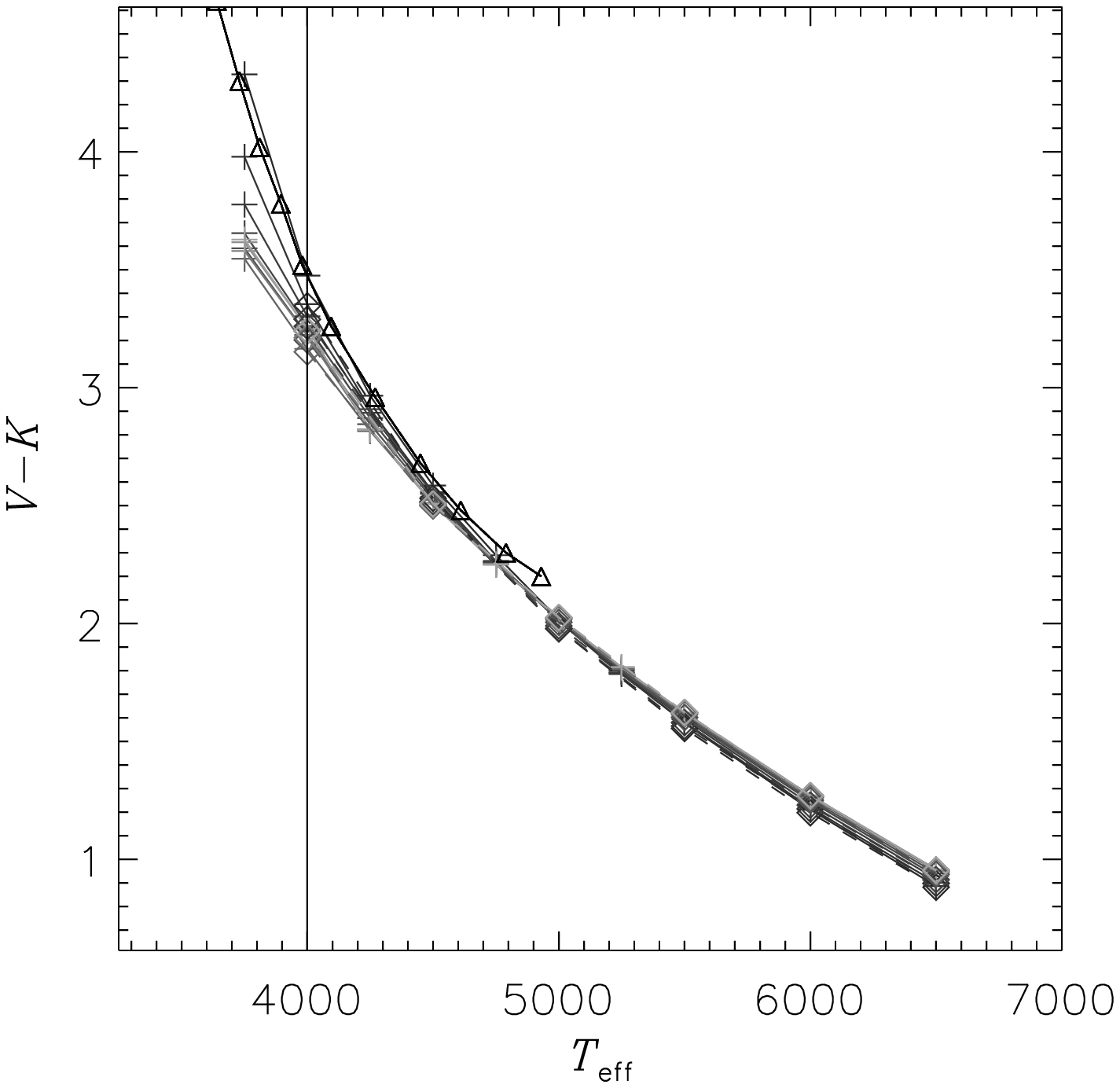}
\caption{$V-K(T_{\rm eff})$ relation.  Synthetic colors from LTE (solid gray-scale lines and crosses) and 
NLTE (dashed gray-scale lines and diamonds), and the calibration of \citet{ridgway80} (solid black line and 
triangles).  The synthetic colors were tied to the \citet{ridgway80} relation with a single-point calibration
of the $[{{\rm M}\over {\rm H}}]=0.0$ models at $T_{\rm eff} = 4000$ K (vertical line).  Gray-scale: Darker line indicates 
larger $[{{\rm M}\over {\rm H}}]$ value throughout model grid range of -6.0 to 0.0.  
Note that we only compute NLTE $I$ values
at a subset of the LTE grid - see text.  
  \label{vmkteff}}
\end{figure}

  Fig. \ref{vmkteff} shows the LTE and NLTE model $V-K(T_{\rm eff})$ relation for our range of model $[{{\rm M}\over{\rm H}}]$
values, over-plotted with the $V-K(T_{\rm eff})$ relation for giants of \citet{ridgway80}.  The  model 
$V-K(T_{\rm eff})$ relation flattens with decreasing $[{{\rm M}\over{\rm H}}]$ value, which is to be expected 
because line blanketing extinction in the $V$ band increases more rapidly with increasing $[{{\rm M}\over{\rm H}}]$
value than does that in the $K$ band.  Within the range of
overlap in $T_{\rm eff}$ (3750 to 5000 K), our model $V-K(T_{\rm eff})$ relation at $[{{\rm M}\over{\rm H}}] = 0.0$
closely tracks that of \citet{ridgway80}, but is slightly steeper.  However, the \citet{ridgway80} sample
of red giants probably includes stars of {\bf $[{{\rm Fe}\over{\rm H}}] < 0.0$}, so it should have a flatter $V-K(T_{\rm eff})$ 
relation than that of $[{{\rm M}\over{\rm H}}] = 0.0$.  NLTE effects are negligible, which is to be expected 
given that these are broad-band colors that average the effects of many spectral lines, and that line blanketing
opacity is already considerably reduced in the $V$ band as compared to the $B$ and $U$ bands.

  \subsection{XMP indices}

  Table \ref{XMPIs} displays, for a selection of $T_{\rm eff}$ values spanning our grid,
the range of $[{{\rm M}\over{\rm H}}]$ values for which each index, $I$,
is a sensitive $[{{\rm M}\over{\rm H}}]$ indicator as judged by the criterion that 
$\Delta[{{\rm M}\over{\rm H}}]\times {{\partial I}\over {\partial [{{\rm M}/{\rm H}}]}} \gtrsim \sigma_{\rm Worthey}$,
where $\Delta[{{\rm M}\over{\rm H}}] \approx 1$ and $\sigma_{\rm Worthey}$ is an observational uncertainty 
described below.  
  Lick indices that meet this criterion and remain strong enough at VMP-to-XMP metallicities 
($[{{\rm M}\over{\rm H}}] < -4.0$) to be detectable are considered to be ``Lick-XMP'' indices.
Generally, {\it all} the ``metallic'' atomic and molecular indices 
have ${{\partial I}\over {\partial [{{\rm M}/{\rm H}}]}}$ values
that increase with decreasing $T_{\rm eff}$ value, and we expect that Lick-XMP indices
will be more readily identifiable at the cool end of our grid.  

\paragraph{}

Table \ref{XMPIs} includes indications
for indices that become ``pathological'' over  a significant $[{{\rm M}\over{\rm H}}]$ range
at {\it any} $T_{\rm eff}$ values by becoming multi-valued, or negative in the case of
those indices that are in linear $W_\lambda$ units.  These pathologies often reflect complications
in the bracketing pseudo-continuum wave-bands as defined by the Lick index standard rather than
with the central feature itself, and most often appear at higher $[{{\rm M}\over{\rm H}}]$ values.

\paragraph{}

  At $T_{\rm eff}=3750$ K we have found five Lick indices that meet
our Lick-XMP criterion all the way down to $[{{\rm M}\over{\rm H}}] = -6.0$ 
(Fe5270, Fe5335, Fe5406, Mg $b$, Na $D$), and four others that meet the criterion 
down to $[{{\rm M}\over{\rm H}}] = -5.0$ (Fe4383, Fe4531, Fe5015, Mg$_{\rm 1}$).
At $T_{\rm eff}=4500$ K there are none that meet the criterion down to $[{{\rm M}\over{\rm H}}] = -6.0$,
and four of the above nine that still meet the criterion down to $[{{\rm M}\over{\rm H}}] = -5.0$ 
(Fe5335, Fe5406, Mg $b$, Na $D$).  These nine Lick-XMP indices are indicated 
in Figs. \ref{specGrid} and \ref{specDiff} with identification labels that
are off-set from the rest.  
By $T_{\rm eff}=5000$ K (and hotter) there are {\it no} indices that 
meet our Lick-XMP criterion.  (Furthermore, we caution that Mg $b$ becomes multi-valued
as a function of $[{{\rm M}\over{\rm H}}]$ for $T_{\rm eff} \gtrsim 5000$ K.)  
The presence of Fe4383, Fe5270 and Fe5335 among our Lick-XMP indices is not surprising
given that \citet{gorgas93} identified them as strong {\bf $[{{\rm Fe}\over{\rm H}}]$-indicators}. 
Na $D$ is an
interesting member of our Lick-XMP indices in that it is not Fe-dominated, but we caution, again, that its usefulness
is compromised by significant ISM extinction.  
As noted in Section \ref{sModel}, our treatment of Mg $b$ and Mg$_{\rm 1}$ is least accurate because we 
neglect $\alpha$-enhancement in the current investigation, and Mg is an $\alpha$-process element.  Therefore,
the following results for these two indices are most suspect and require further investigation.

\paragraph{}

\begin{figure}
\plotone{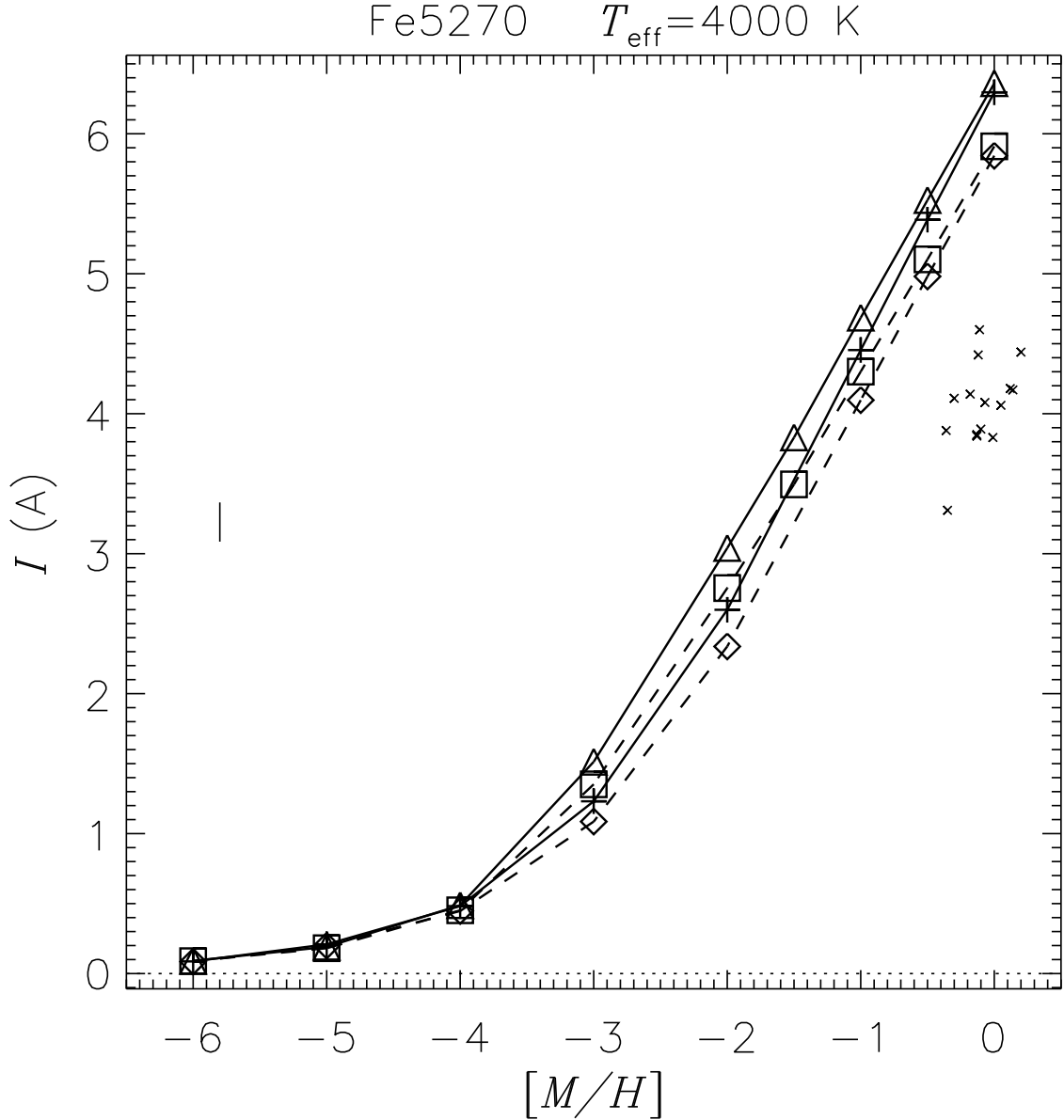}
\caption{Index Fe5270, one of our identified Lick-XMP indices: $I$ as a function of 
$[{{\rm M}\over {\rm H}}]$ at $T_{\rm eff} = 4000$ K.  Vertical
line in middle left: Indication of the observational uncertainty, $\sigma_{\rm Worthey}$, of W94 - see text.  
$I$ has been computed from spectra with $R$ values of SDSS (solid line) and IDS (dashed line)
spectroscopy that were
computed in LTE (triangles, squares) and NLTE (crosses, diamonds) - see text.  The vertical spread in 
$I$ values at a given $[{{\rm M}\over {\rm H}}]$ value can be taken as an 
estimate of ``spectroscopic and modeling physics uncertainty''.
Note that we only compute NLTE $I$ values at a subset of the LTE grid - see text.  
For comparison we show the observationally derived $I$ and 
$T_{\rm eff}$ values of \citet{wortheyo97} for giants ($1.0 \leq \log g \leq 3.0$) of $T_{\rm eff}$ 
within $\pm 100$ K (``X'' symbols). 
No calibration of our $I$ values to those of \citet{wortheyo97} has been performed.
\label{Fe5270_4000}}
\end{figure}

\begin{figure}
\plotone{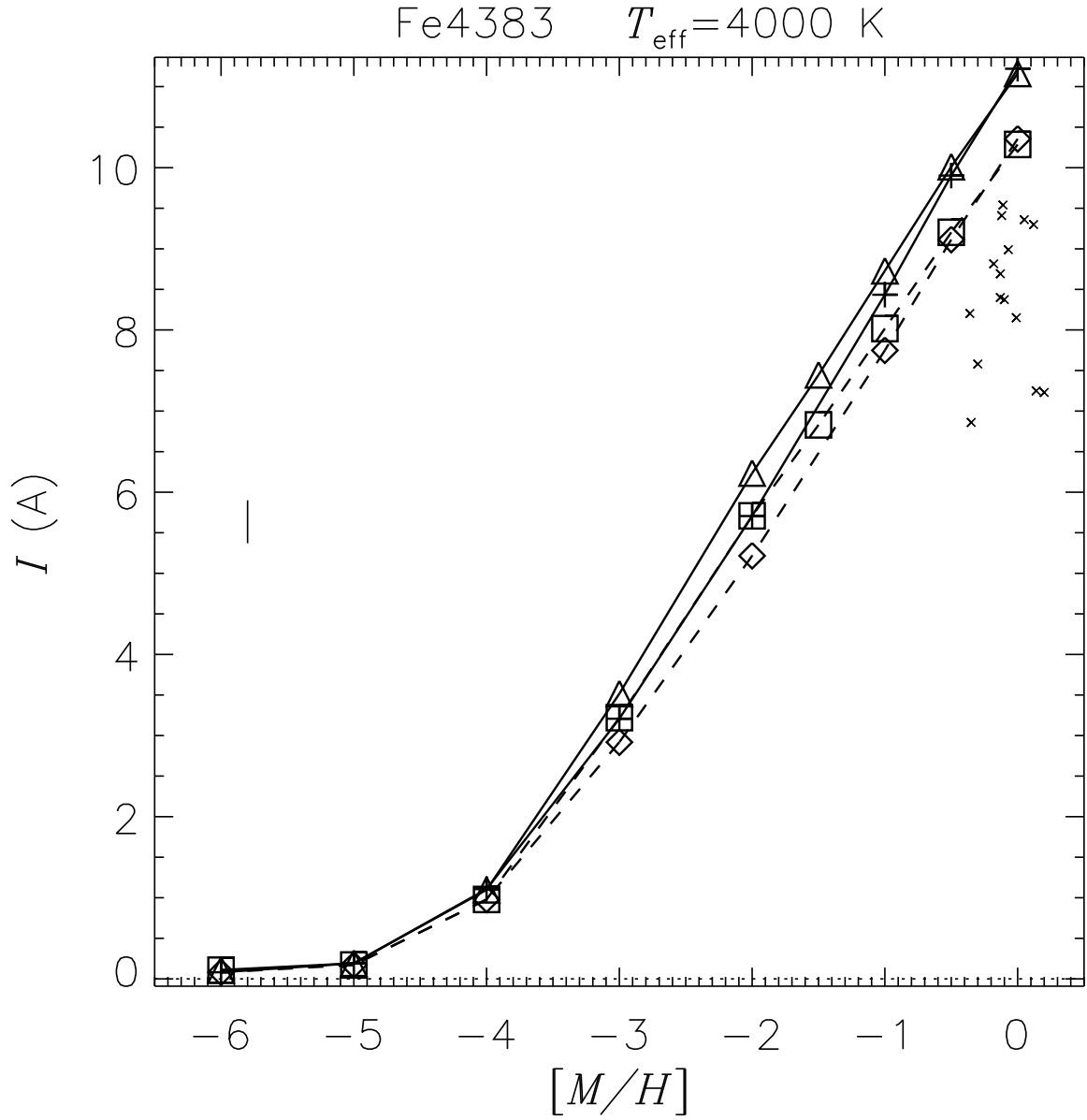}
\caption{Same as Fig. \ref{Fe5270_4000}, but for index Fe4383.
\label{Fe4383_4000}}
\end{figure}

\begin{figure}
\plotone{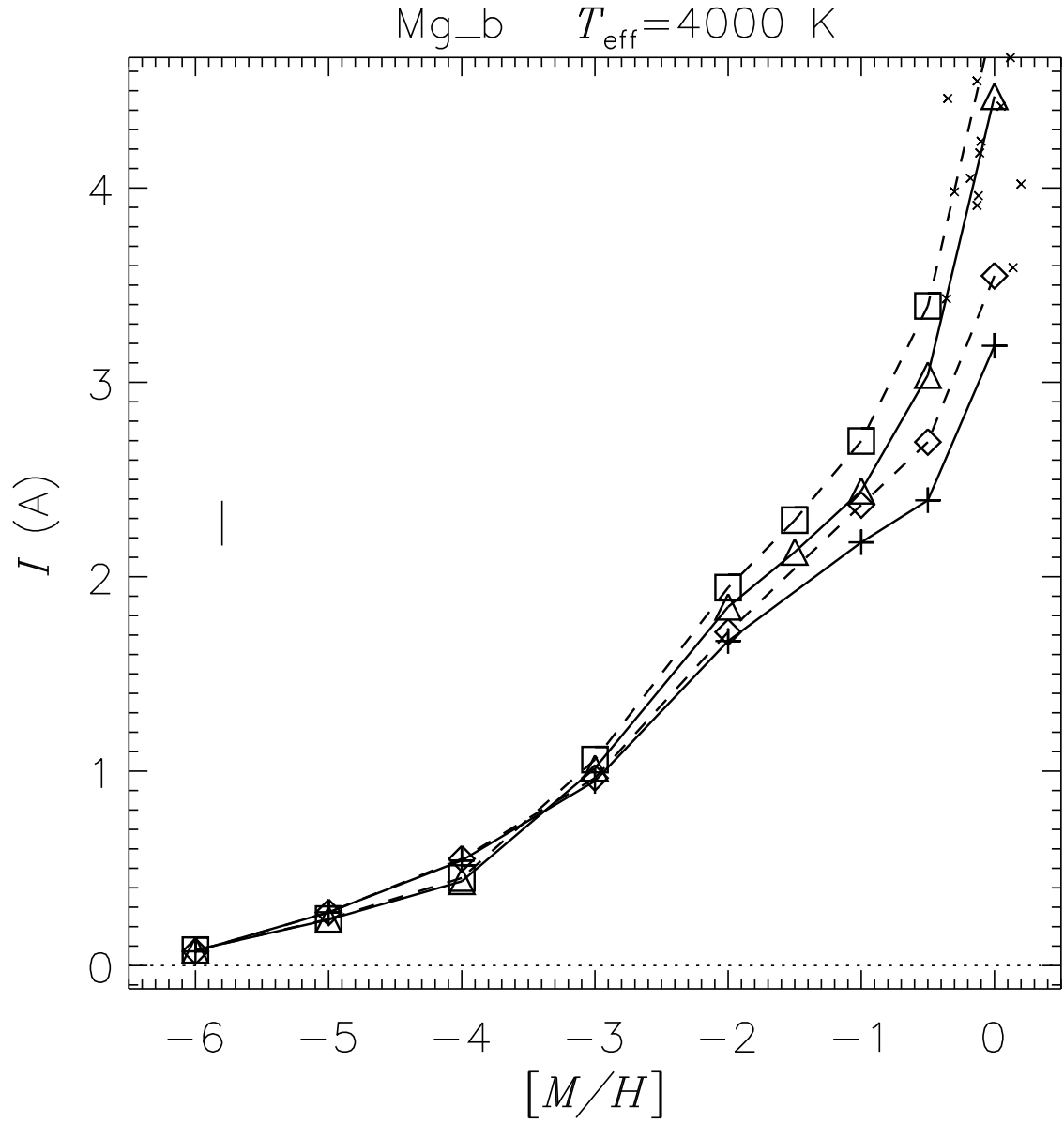}
\caption{Same as Fig. \ref{Fe5270_4000}, but for index Mg $b$, an example of one of our identified Lick-XMP 
indices, {\it not} dominated by Fe.
\label{Mgb_4000}}
\end{figure}

\begin{figure}
\plotone{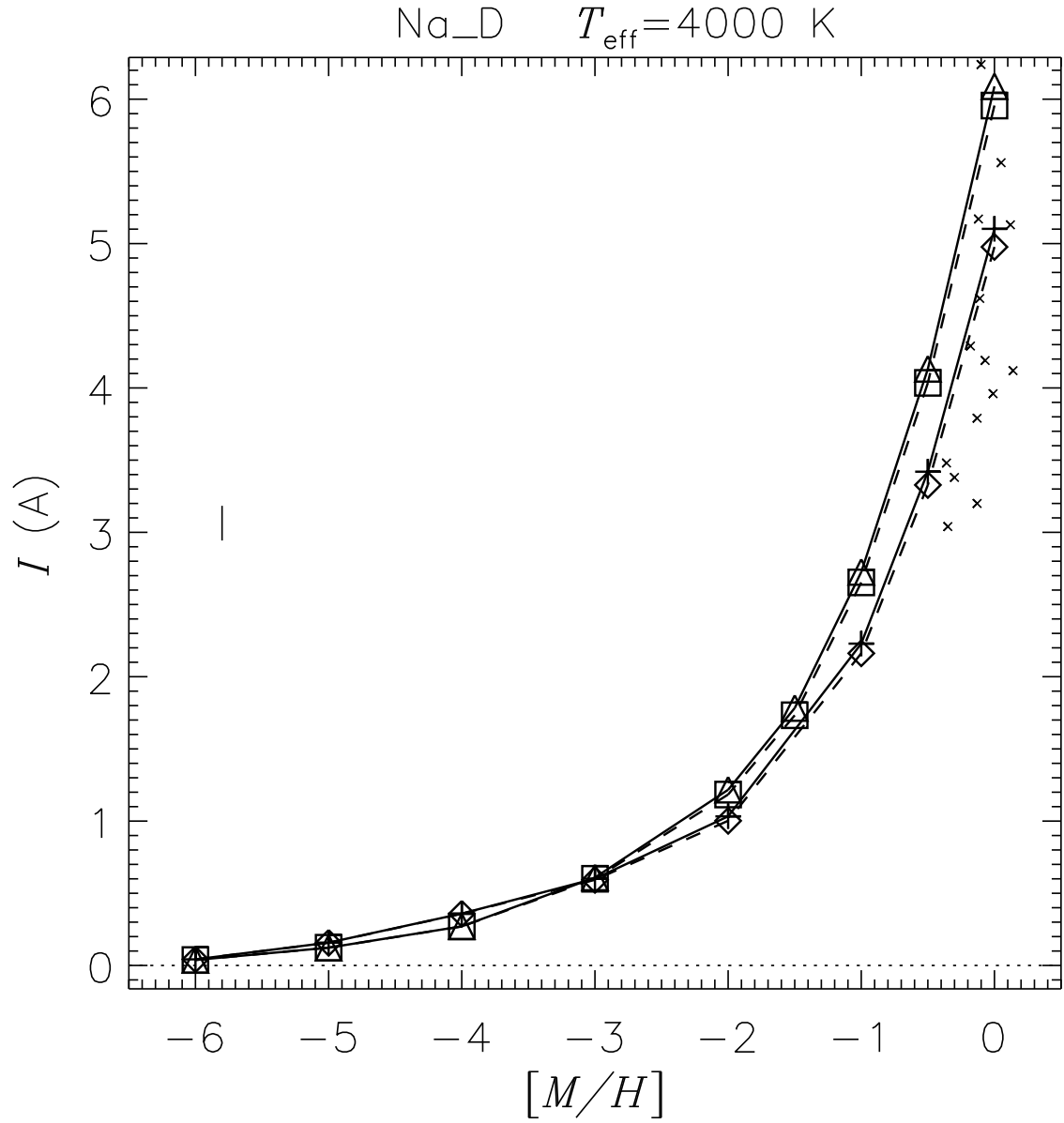}
\caption{Same as Fig. \ref{Fe5270_4000}, but for index Na D, another example of one of our identified Lick-XMP 
indices, {\it not} dominated by Fe.
\label{NaD_4000}}
\end{figure}

Figs. \ref{Fe5270_4000} to \ref{NaD_4000} show the modeled $I([{{\rm M}\over{\rm H}}])$ relation 
at $T_{\rm eff}=4000$ K, and Figs. \ref{Fe5270_0} to \ref{NaD_0} show the modeled $I(T_{\rm eff})$ relation at
$[{{\rm M}\over{\rm H}}]=0.0$ for four of our Lick-XMP indices, Fe5270, Fe4383, Mg $b$, and Na $D$.  The 
latter are the only two Lick-XMP indices {\it not} dominated by Fe. 
Results are shown for spectra computed with the
$R$ values of SDSS and IDS spectroscopy, as computed in LTE and NLTE. 
For comparison, we have over-plotted observationally derived $I(T_{\rm eff})$ 
and $I([{{\rm M}\over{\rm H}}])$ points for a wide range of cool giants from the catalog of  \citet{wortheyo97} with catalog $\log g$ values between 1.0 and 3.0, and $T_{\rm eff}$ and $[{{\rm M}\over{\rm H}}]$ within $\pm 100$ K and $\pm 1.0$, 
respectively, of the plotted models.  Note that the \citet{wortheyo97} $I$ data 
only includes stars of $[{{\rm M}\over{\rm H}}] \gtrsim -1.0$, and that {\it no} 
calibration of our $I$ values to those of \citet{wortheyo97} has been performed. 
Nevertheless, from Figs. \ref{Mgb_4000} and \ref{NaD_4000}, the agreement between our modeled 
Mg $b$ and Na $D$ $I$ values and the measurements of \citet{wortheyo97} for stars
of $T_{\rm eff} = 4000 \pm 100$ K is assuring.
Figs. \ref{Fe5270_4000} to \ref{NaD_4000} show that these indices generally satisfy our 
Lick-XMP criteria for $T_{\rm eff}$ values in the cool part of our grid.  
Fig. \ref{Fe4383_0} shows that the utility of Fe4383 is somewhat compromised - it is double-valued as a function
of $T_{\rm eff}$ at the cool edge of our grid.  However, it is a sensitive and
useful $[{{\rm M}\over{\rm H}}]$ discriminator otherwise, and we include it because 
there are few indices that qualify as Lick-XMP at all.
Figs.  \ref{Fe5270_0} to \ref{NaD_0} show that, as discussed above, Lick-XMP indices
quickly become weaker and lose their ability to discriminate among $[{{\rm M}\over{\rm H}}]$
values for $T_{\rm eff} \gtrsim 5000$ K.  Figs. \ref{Fe5270_4000} through \ref{NaD_0}
also include an indication of the ``observational uncertainty'' as determined by W94,
$\sigma_{\rm Worthey}$ (see below), to aid in assessing the significance of $\Delta I$ differences. 

\paragraph{}

The effect of NLTE is complex in that every Lick index is really a compound feature 
caused by significant spectral lines from several species.  Moreover, although
the bracketing pseudo-continua used to define the indices were chosen to be relatively 
insensitive to stellar parameter values, NLTE effects on line strengths in the
bracketing regions will, in principle, also play a role in the overall effect of NLTE
on the computed $I$ value.  However, Fe5270 is typical
of our results for the Fe-dominated indices in that the effect of the
well-known NLTE over-ionization of \ion{Fe}{1} in late-type stars \citep{rutten86}
leads to smaller $I$ values at every $T_{\rm eff} - [{{\rm M}\over{\rm H}}]$
combination.  Figs. \ref{Fe5270diff} and \ref{Mgbdiff} show the size of the
NLTE effect, $\Delta I\equiv I_{\rm NLTE} - I_{\rm LTE}$, as a function of
$T_{\rm eff}$ for Fe5270 and Mg $b$.  The effect of NTLE on strong low $\chi$ \ion{Fe}{1} lines 
is to weaken them (a negative correction to {\it modeled} $I$ value) as a result of NLTE over-ionization.  
In late-type stars the effect will generally be maximal where 
the discrepancy, $\Delta T$, throughout the line forming region between the radiation temperature, $T_{\rm Rad}$, of 
the photo-ionizing near-UV band radiation in the NLTE treatment,
and the local kinetic temperature, $T_{\rm Kin}$, that determines the ionization
balance in the LTE treatment, is largest.  For \ion{Fe}{1} in giants, the discrepancy is largest
around $T_{\rm eff}\approx 5000$ K, and decreases in magnitude for both lower and higher $T_{\rm eff}$ value  
(see \citet{rutten86} for a thorough analysis for the case of Fe in the Sun). 

\paragraph{}

For \ion{Mg}{1} (Index Mg $b$) the NLTE correction to the modeled $I$ value is also 
negative, but increase in magnitude with decreasing $T_{\rm eff}$ throughout this $T_{\rm eff}$ range,
especially for $T_{\rm eff} < 4000$ K.  \citet{osorio15} very recently conducted a thorough  
NLTE analysis of \ion{Mg}{1} in late-type stellar atmospheres, including an investigation
of \ion{H}{1} collisional cross-sections and electron-exchange reactions.  For the $\lambda$ 5184 line,
they found that NLTE effects lead to a weakening of the modeled line
(hence the {\it positive} abundance correction in their Fig. 10), by an amount that depends
on choice of atomic data, but can be as much as 0.4 dex at $T_{\rm eff}=4500$ K, $\log g=1.0$,
and $[{{\rm M}\over{\rm H}}]=0.0$, which is qualitatively consistent with our result. 


\paragraph{}

Interestingly, the {\it magnitude} of NLTE effect on the computed $I$ value is comparable to that caused by 
 changing spectral resolution ($R$ value of IDS $vs$ SDSS).  Na $D$ is an exception because it 
is so broad that it
is minimally affected by the choice of $R$. 
Computing $I$ from higher $R$-value spectral ({\it i.e.} that of SDSS) can either increase
or decrease the $I$ value, depending on the index.  The same remarks as made when
considering NLTE effects above also apply: namely, that the effect of $R$ value on any given
index will depend on how the bracketing pseudo-continua are affected as well as the 
central feature itself. 
Altogether, the vertical spread in $I$ values
at each abscissa can be taken as an approximate indication of ``spectroscopic and modeling
physics uncertainty''.

\subsection{Polynomial fits \label{sfitfunc}}

  Based on the approach taken by  W94 with {\it observed} IDS spectra {\bf and observationally determined stellar parameters}, 
we have used multi-variate linear regression to determine ten polynomial fitting
(regression) coefficients (or model parameters), $\{ C_{\rm n} \}$, for each index, $I$, for a polynomial fitting function, $P_{\rm 3}$, that accounts for all terms,
including cross-products, up to third order in $[{{\rm M}\over{\rm H}}]$ and $\log\theta$, where $\theta\equiv {\bf 5040}/T_{\rm eff}$, 

\begin{eqnarray}
 I \approx P_{\rm 3} \equiv C_{\rm 0} + C_{\rm 1}[{{\rm M}\over{\rm H}}] + C_{\rm 2}\log\theta  
 + C_{\rm 3}[{{\rm M}\over{\rm H}}]^2 + C_{\rm 4}\log^2\theta + C_{\rm 5}[{{\rm M}\over{\rm H}}]\log\theta \nonumber\\
 + C_{\rm 6}[{{\rm M}\over{\rm H}}]^3 + C_{\rm 7}\log^3\theta 
 + C_{\rm 8}[{{\rm M}\over{\rm H}}]^2\log\theta + C_{\rm 9}[{{\rm M}\over{\rm H}}]\log^2\theta 
\label{poly}
\end{eqnarray}

 Note that, as per convention, the units of $I$ for the molecular indices CN$_{\rm 1}$, CN$_{\rm 2}$, Mg$_{\rm 1}$, 
Mg$_{\rm 2}$, TiO$_{\rm 1}$ and TiO$_{\rm 2}$ are magnitudes, and those for the remaining indices are \AA.
Moreover, following W94, in the special case of the TiO$_{\rm 1}$ and TiO$_{\rm 2}$ indices, which exhibit
a rapid increase in strength with decreasing $T_{\rm eff}$ near the lower limit of our $T_{\rm eff}$
range, we fit
Eq. \ref{poly} to $\log I$ so that $I$ is being fit by $\exp P_{\rm 3}$.  
The fitting is performed with the intrinsic REGRESS procedure in Interactive Data Language (IDL) and is 
achieved by minimizing
the $\chi^2$ figure of merit assuming that the uncertainty, $\sigma_{\rm i}(I)$, of all ``data'' points 
$I_{\rm i}$, is unity.  We note that IDL installations include the source code for all intrinsic procedures,
and we have been able to critically inspect the REGRESS procedure.  We use REGRESS to compute nine linear 
regression coefficients and a constant term in Equation \ref{poly} for nine basis functions that consist of the powers of 
the independent parameters ($T_{\rm eff}$, $[{{\rm M}\over{\rm H}}]$) and their products up to third order.
This amounts to fitting a model with nine parameters to 90 data points (90 computed $I$ values for ten $T_{\rm eff}$
and nine $[{{\rm M}\over {\rm H}}]$ values), thus having 81 degrees of freedom.
This is consistent with the fitting method of W94 - we note that their Equation 4 appears to be the
standard formula for $\chi^2$ because their summed square deviations are weighted by their inverse observational
uncertainty, $1/\sigma^2$, although they have labeled their figure of merit ``rms$^2$''.  We note
that because we have no proper data uncertainties ({\it i.e.} ``measurement errors'' in data modeling), 
$\sigma_{\rm i}(I)$, we cannot properly propagate 
errors to compute uncertainties for the fitted model parameters, $C_{\rm n}$ - rather we compute ``fitting uncertainties'',
$\sigma$, {\it post hoc} from the calculated value of $\chi^2$ and the number of degrees of freedom. 
Assuming these ``fitting uncertainties'' reflect errors that are normally distributed,
they may be interpreted as $68\%$ confidence intervals for the fitted values of the corresponding $C_{\rm n}$.     

\paragraph{}

Tables \ref{coeffsthetaIDS} and \ref{coeffsthetaSDSS} present these $\{ C_{\rm n} \}$ values for LTE spectra of IDS and 
SDSS resolution, respectively,
for the nine indices that were identified above as good Lick-XMP diagnostics,
in the same format and numerical precision as that of Table 2 of W94 (``Data for stars of 
$3570 < T_{\rm eff} < 5160$ K'') for direct comparison.  Table \ref{coeffsvmkIDS} presents the results for
IDS spectral resolution of 
performing the same third order multiple linear regression
with the model $V-K$ color in place of $\log\theta$.  We also present the values of the standard
deviations, $\sigma$, computed for each $C_{\rm n}$ parameter from $\chi^2$, and the value of the
reduced $\chi^2$ given 81 degrees of freedom, although we caution that in the absence of proper
measurement errors, $\chi^2$ is not really a goodness-of-fit figure of merit.
Figs. \ref{residTeffFe5270} through \ref{residTeffNaD} show the comparison of the polynomial fits
to the modeled $I(T_{\rm eff})$ relation for $[{{\rm M}\over {\rm H}}]=0.0$ at SDSS and IDS 
resolution, and the residuals.  For SDSS resolution, we also show both the fitted relation and
the residuals computed with the 1-$\sigma$ ``fitting uncertainties'' (see above) added and subtracted 
from each of the $C_{\rm n}$ values to illustrate these two limiting cases.  Figs. \ref{residAbndFe5270}
through \ref{residAbndNaD} present the same information for the polynomial fits
to the modeled $I([{{\rm M}\over {\rm H}}])$ relation for $T_{\rm eff}=4000$ K.

\paragraph{}

  There are a number of important differences between our approach and that of W94:\\

 a) Our $C_{\rm n}$ values complement those of W94 by being based on spectra of SDSS resolution
rather than those of IDS resolution, and are more relevant to both SDSS and LAMOST spectra
and to the investigation of \citet{franchini10}. \\

 b) $\log g$ is not a fitting parameter, so we do not have cross-product terms that capture the
dependence of $I$ on the product of $\log g$ or any of its powers and other parameters or their powers. \\ 

 c) Table 2 of W94 contains coefficients
for fits in the $T_{\rm eff}$ range of 3570 to 5160 K, whereas our fits apply to the range 3750 to 6500 K.  The
difference at the high $T_{\rm eff}$ end is necessary for us to accommodate the $T_{\rm eff}$ range of interest for detected halo red giants. 
Because the lower limit of our $T_{\rm eff}$ range is significantly higher than that of W94, results for our
 TiO indices are especially suspect, and not comparable to W94.
W94 included stars of $T_{\rm eff} > 5160$ K in their fits for warm and hot stars (5040 to 13260 K, their Table 3).
This is well beyond the limits of our red giant grid and we are not able to compare to their ``hot star'' fits. \\

d) W94 carry out a careful statistical $F$-value test of the goodness-of-fit to separately determine whether the addition of
each successive 
$C_{\rm n}$ term in Eq. \ref{poly} led to a statistically significant change in the fitted $I$ value for each index.  As a result,
many of the $C_{\rm n}$ values in their Table 2 are blank because, presumably, including the corresponding term
in the fitted function led to an insignificant reduction in the variance.  We have chosen to simply let the 
regression find
$C_{\rm n}$ values for all ten terms in Eq. \ref{poly} consistently for all indices, with the expectation that
terms that are of low significance will have small fitted values of the corresponding $C_{\rm n}$.  Our expectation is that a third
order fit is of low enough order that we do not expect high order spurious solutions to compromise the fit, and our situation
is simpler than that of W94 in that $\log g$ is not a fitting parameter.

  \paragraph{Error analysis }

  W94 describes the uncertainty in determining an $I$ value as a ``typical rms error per 
observation'', which we denote $\sigma_{\rm Worthey}$.  For reference, we have included an indicator of the magnitude 
of $\sigma_{\rm Worthey}$ for
each index in Figs. \ref{Fe5270_4000} through \ref{NaD_0}.  
W94 quantifies the uncertainties in the fitted $I$ values with a ``residual'' rms value 
in units of the observational uncertainty, $\sigma_{\rm Worthey}$.  
We are not working with observational data, and we quantify the 
uncertainties in our fits, $\sigma$, with the quadrature sum of the 1-$\sigma$ uncertainty estimates computed 
for each $C_{\rm n}$ value from the multiple linear regression
procedure (as described above), and have included them in 
Tables \ref{coeffsthetaIDS} through \ref{coeffsvmkIDS}. 
Generally, the first order coefficients, $C_{\rm 1}$ for $[{{\rm M}\over{\rm H}}]$ and $C_{\rm 2}$ for
$T_{\rm eff}$, from the fit to SDSS resolution spectra are larger than those form the fit to the IDS
resolution spectra.  This is to be expected because in spectra of higher $R$ value the first order dependence
of the strength of spectral features on stellar parameters is less diluted by the re-shuffling of information
within the instrumental spectral profile. 

\paragraph{}
 
The coefficient of the $\log ^3\theta$ term, $C_{\rm 7}$, generally had, by far, the largest
1-$\sigma$ uncertainty value of all the $C_{\rm n}$ values, and dominates our quadrature
sum $\sigma$ values.  Generally, the parameter with next largest 1-$\sigma$ uncertainty
was the $\log ^2\theta$ term, $C_{\rm 4}$, but it was much smaller than that of $C_{\rm 7}$. 
Table \ref{coeffsthetaIDS} shows that the magnitude of $C_{\rm 7}$ is generally larger than
any of the other coefficients, and this is consistent with what was reported in Table 2 of 
W94.  We also found that the value of $C_{\rm 7}$ was the most sensitive to spectral 
resolution, differing by as much as a factor of four between the fits to IDS and SDSS resolution
spectra.  We conclude that the terms in non-linear powers of $\log\theta$ were generally least well fit. 
However, for these GK stars, $\theta\equiv 5040/T_{\rm eff}$ is of order unity, and
the squares and cubes of the independent variable $\log\theta$ is much less than unity.  We have found
that the 1-$\sigma$ uncertainty of the $C_{\rm 4}$ and $C_{\rm 7}$ terms for the analogous $V-K^{\rm 2}$ and 
$V-K^{\rm 3}$ terms from the fits with $V-K$ in lieu of $T_{\rm eff}$ are much smaller 
and are consistent with those of the other $C_{\rm n}$ coefficients, and from Table \ref{coeffsvmkIDS}
it can be seen that the corresponding total $\sigma$ values are smaller.  

\paragraph{}

  For our nine XMP indices, we compare our fitted $C_{\rm n}$ values for the zeroth 
and first order terms of the fit to IDS resolution spectra with those of W94,
given the four caveats listed above.  For the fits with $\log\theta$ as an independent parameter,
comparable to Table 2 of W94, for five of six indices designated ``Fe'', our fitted $C_{\rm 0}$
value is consistently smaller than that of W94, and ranges from about 0.6 to 0.75 of the W94 value,
and for Fe5270 we are in very close agreement.
For Na $D$, our $C_{\rm 0}$ value is also smaller, but is in closer agreement with that of W94.
Mg$_{\rm 1}$ is our only index for which our $C_{\rm 0}$ is larger than W94, by about a factor of 1.5. 
For Mg $b$ we
find a negative value of $C_{\rm 0}$ where W94 finds a positive value.  In both cases, the magnitude of 
$C_{\rm 0}$ is about unity.  
For the $\log\theta$ coefficient, $C_{\rm 2}$, our values for Fe4531, Fe5335, and Fe5406 are
close to the values of W94.   
The remaining $C_{\rm 2}$ values are also generally within a 
factor of two, greater or less than, and of the same sign as those of W94.  The two exceptions are Na $D$ for 
which our $C_{\rm 2}$ value is about three times larger, and Mg$_{\rm 1}$ for which our $C_{\rm 2}$ value is positive
and that of W94 is negative, although the magnitudes are with a factor of two.  
For the $[{{\rm M}\over{\rm H}}]$ coefficient, $C_{\rm 1}$, our values for Fe5335, Fe5406, Mg $b$, and Mg$_{\rm 1}$ agree closely with
those of W94.  For the remaining XMP-Lick indices, including Na $D$, our $C_{\rm 1}$ values differ by as much as a factor of 2.5,
greater or less then, those of W94.   

   \paragraph{}

 For the fits with $V-K$ as an independent parameter (in lieu of $\log\theta$), only four of our nine 
Lick-XMP indices appear in the comparable Table 4 of W94 (Fe4383, Fe4531, Fe5015, and Fe5406).  
 For Fe5015, W94 does not present a $C_{\rm 0}$ value, but our $C_{\rm 2}$ ($I(V-K)$) agrees
closely with theirs.  For Fe4531, W94 does not present a $C_{\rm 2}$ value, and our $C_{\rm 0}$ value
is about a factor of two larger.  For both of these indices, our $C_{\rm 1}$ values ($I([{{\rm M}\over{\rm H}}])$,
at constant $V-K$ this time) are close to those of W94. 
For Fe4383 and Fe5406
the situation is disconcerting and puzzling - we find both $C_{\rm 0}$ and $C_{\rm 2}$ values with the opposite sign and 
a difference in magnitude ranging from two to four.
For both of these indices, W94 have no $C_{\rm 1}$ value for the term in  $[{{\rm M}\over{\rm H}}]$,
indicating, presumably, they found insignificant {\it linear} dependence of $I$ on $[{{\rm M}\over{\rm H}}]$
at fixed $V-K$,
and, consistently, we find modest $C_{\rm 1}$ values of -0.4489 and -0.0986, respectively.   
We have been unable to identify the
reason for the gross discrepancy in fitted polynomial coefficients for Fe4383 and Fe5406, beyond those
expressed in the caveats itemized above.

\subsection{Special indices } 

\paragraph{CaHK }

 \citet{severn05} introduced a new Lick-type index, which they designate CaHK, and
\citet{franchini10} modified the definition by changing the blue 
pseudo-continuum band-pass and removing the central 5 \AA~ around the line cores to
remove potential chromospheric emission.  \citet{franchini10} found that CaHK had the advantage
of being sensitive to $\alpha$-enhancement, as well as being less influenced by Fe than most of the
other atomic indices.  The index seems to shows promise as a Lick-XMP
$[{{\rm M}\over{\rm H}}]$ diagnostic because the line is very strong.  We do not remove the central 5 \AA~ 
from our index computation because our models have outer atmospheres that are in 
radiative equilibrium and do not have chromospheric emission.  We have confirmed that 
we can reproduce the qualitative double-valued behavior of $I(\theta)$ exhibited in
Fig. 7 of \citet{franchini10}.  We have found that the index is a good Lick-XMP
diagnostic for our models at the cool edge of our grid, of $T_{\rm eff}$ value equal
to 3750 to 4000 K.  However, unfortunately, CaHK is either double-valued
as a function of $[{{\rm M}\over{\rm H}}]$ for $T_{\rm eff}$ in the range of 4250 to 5500 K,
or becomes negligible and insensitive to $[{{\rm M}\over{\rm H}}]$ for $[{{\rm M}\over{\rm H}}] \leq -4$
in the $T_{\rm eff}$ range of 5500 to 6500 K.  We can only recommend CaHK as a
Lick-XMP index for $T_{\rm eff} \leq 4000$ K.  Because the behavior of CaHK is double
valued as a function of both $T_{\rm eff}$ and $[{{\rm M}\over{\rm H}}]$ throughout
much of the $T_{\rm eff} - [{{\rm M}\over{\rm H}}]$ plane, the $C_{\rm n}$ and partial
derivative values are probably not useful for interpolation, and we have not included them.
 
\paragraph{TiO }

  We have included the TiO$_{\rm 1}$ and TiO$_{\rm 2}$ indices in our analysis because
they are part of the Lick system, but they are only significant in strength and 
in $[{{\rm M}\over{\rm H}}]$-sensitivity for the coolest part of our grid, and then
only at the highest $[{{\rm M}\over{\rm H}}]$ values.  TiO$_{\rm 2}$ is the stronger
of the two indices and remains sensitive to $[{{\rm M}\over{\rm H}}]$ in the -2.5 to 0.0
range to $T_{\rm eff}$ values as high as 4250 K.  Neither TiO index has pathologies
such as being double-valued.
Note that we fit Eq. \ref{poly} to $\log I$ to
account for the strong $T_{\rm eff}$ and $[{{\rm M}\over{\rm H}}]$ dependence of 
TiO$_{\rm 1}$ and TiO$_{\rm 2}$, following W94.  Because our grid does not extend to
the low $T_{\rm eff}$ values included in the W94 fit, where TiO is strong, our $C_{\rm n}$ and
partial derivative values are not comparable to W94, and should be treated with more caution
that those of the other indices, and we do not include them here.

\subsection{Partial derivatives with respect to $T_{\rm eff}$, $V-K$ and $[{{\rm M}\over{\rm H}}]$ }

  We have used our $\{ C_{\rm n} \}$ values to compute the partial derivatives 
$100~ {\rm K}\times {{\partial I}\over{\partial T_{\rm eff}}}|_{[{{\rm M}/{\rm H}}]}$, 
$0.5\times {{\partial I}\over{\partial [{{\rm M}/{\rm H}}]}}|_{T_{\rm eff}}$,
$0.25~ {\rm mag}\times {{\partial I}\over{\partial (V-K)}}|_{[{{\rm M}/{\rm H}}]}$,
and $0.5\times {{\partial I}\over{\partial [{{\rm M}/{\rm H}}]}}|_{(V-K)}$
 for all Lick indices modeled in LTE, 
in both index units and in units of $\sigma$, as defined above, for IDS and SDSS resolution spectra.
For example, the partial derivative with respect to $\log\theta$ at constant $[{{\rm M}/{\rm H}}]$ can
be found from

\begin{eqnarray}
{{\partial I}\over{\partial\log\theta}}|_{[{\rm M}/{\rm H}]} = 
C_{\rm 2} + 2 C_{\rm 4}\log\theta + C_{\rm 5} [{{\rm M}\over {\rm H}}] +  \nonumber\\
3 C_{\rm 7}\log ^2\theta + C_{\rm 8} [{{\rm M}\over {\rm H}}]^2 + 
2 C_{\rm 9}\log\theta [{{\rm M}\over {\rm H}}] .
\end{eqnarray}

Then, the partial derivative with respect to $T_{\rm eff}$ follows from

\begin{equation}
{{\partial I}\over{\partial T_{\rm eff}}} = - {{\partial I}\over{\partial\log\theta}} \log e / T_{\rm eff} .
\end{equation}

Tables \ref{partialteffs} and \ref{partialvmks} present the values for our nine identified
Lick-XMP indices for SDSS resolution only, and are comparable
to Tables 7A and 7B of W94.
The values in Table \ref{partialteffs}, 
provide an indication of by how much
the measured value of $I$ will differ between two stars that differ by $\Delta T_{\rm eff}\approx 100$ K at each
value of $[{{\rm M}\over{\rm H}}]$, and 
by how much $I$ will differ between two stars that differ by 
$\Delta [{{\rm M}\over{\rm H}}]\approx 0.5$ at each value of $T_{\rm eff}$.  
Alternately, these derivatives can be used to estimate the change in inferred 
$T_{\rm eff}$ or $[{{\rm M}\over{\rm H}}]$ as a result of the change in computed $I$ caused
by NLTE effects ({\it i.e.} where $\Delta I \equiv I_{\rm NLTE} - I_{\rm LTE}$). 
The partial derivative values 
that are in units of $\sigma$ give
an indication of the significance, or detectability, of these changes in $I$ value given the formal
uncertainty $\Delta I \approx\sigma$. 

\paragraph{} 

 As an example of the implications of NLTE effects for $[{{\rm M}\over {\rm H}}]$
determination, from Table \ref{partialteffs}, for Fe5270 the quantity 
$0.5\times {{\partial I}\over{\partial [{{\rm M}/{\rm H}}]}}|_{T_{\rm eff}} = 1.2717$
at $(T_{\rm eff}, [{{\rm M}/{\rm H}}]) = (5000~ {\rm K}, 0.0)$, and from Fig.
\ref{Fe5270diff}, the computed change in $I$ caused by NLTE effects, $\Delta I$,
at 5000 K is {\bf $\sim -0.05$}.  
{\bf This corresponds to an LTE model $[{{\rm M}/{\rm H}}]$ value that is smaller by}

\begin{equation}
\Delta [{{\rm M}/{\rm H}}] \approx 0.5\times -0.05 / 1.2717 \approx -0.02 
\end{equation}  

{\bf Inversely, fitting a given observed $I$ value with NLTE as compared to LTE models would require a
compensating model value of $[{{\rm M}/{\rm H}}]$ that is $\sim 0.02$ larger,
consistent with the sign of the change in inferred $[{{\rm M}\over {\rm H}}]$ at fixed
$I$ value for $T_{\rm eff}=4000$ K shown in Fig. \ref{Fe5270_4000}.}
We emphasize that this is an estimate of $\Delta [{{\rm M}/{\rm H}}]$ based on the
modeled {\it LTE} value of ${{\partial I}\over{\partial [{{\rm M}/{\rm H}}]}}|_{T_{\rm eff}}$.
A more accurate estimate would follow from a NLTE value of 
${{\partial I}\over{\partial [{{\rm M}/{\rm H}}]}}|_{T_{\rm eff}}$, and the importance of this
consideration depends on the magnitude of the difference between 
${{\partial I}\over{\partial [{{\rm M}/{\rm H}}]}}|_{T_{\rm eff}}$ values computed from NLTE and
LTE model grids.  However, computation of
NLTE partial derivatives that are comparable to those of LTE requires a NLTE model grid that
includes {\it all} the same $(T_{\rm eff}, [{{\rm M}/{\rm H}}])$ points as the LTE grid.
Currently, our NLTE model grid covers only a subset because of the larger computational cost
of NLTE modeling, and its only purpose is to spot check the effect of NLTE on computed $I$
values at select grid point. 

\section{Conclusions \label{sConc}}

Using a grid of red giant synthetic spectra that extends from solar- to XMP-metallicity 
($[{{\rm M}\over {\rm H}}]=-6.0$) we have identified nine of the original 21 Lick indices,
designated Lick-XMP indices, 
that remain significantly detectable and significantly sensitive to $[{{\rm M}\over {\rm H}}]$
down to XMP values (at least $[{{\rm M}\over {\rm H}}]=-5.0$) for giants of 
$T_{\rm eff} < 4500$ K.  For warmer late-type giants,
{\it all} Lick indices become undetectable or insignificantly sensitive to $[{{\rm M}\over {\rm H}}]$
before $[{{\rm M}\over {\rm H}}]$ decreases to -4.0.  The Lick-XMP indices should be the most useful ones for 
characterizing very old ``fossil'' stars that formed very early in the history of the Galaxy,
and in other galaxies.  We also investigated a newer Lick-type index, CaHK, introduced by
\citet{severn05} and developed by \citet{franchini10} as a potential Lick-XMP index, given
its strength.  However, for CaHK, $I$ is double valued as a function of $[{{\rm M}\over {\rm H}}]$
and $T_{\rm eff}$ over much of our grid and its usefulness is restricted to the cool edge of our grid
($T_{\rm eff} < 4000$ K).

\paragraph{}

  For our LTE grid of SDSS resolution spectra we present polynomial coefficients, $\{C_{\rm n}\}$, 
to third order in the independent variable pairs ($\log\theta, [{{\rm M}\over {\rm H}}]$) 
and ($(V-K), [{{\rm M}\over {\rm H}}]$) derived from multi-variate linear regression, approximately 
comparable to those of W94 for IDS
resolution spectra.  We present the partial derivatives  
${{\partial I}\over {\partial T_{\rm eff}}}|_{[{{\rm M}/{\rm H}}]}$,
${{\partial I}\over {\partial [{{\rm M}/{\rm H}}]}}|_{T{\rm eff}}$,
${{\partial I}\over {\partial (V-K)}}|_{[{{\rm M}/{\rm H}}]}$,
and ${{\partial I}\over {\partial [{{\rm M}/{\rm H}}]}}|_{(V-K)}$
computed from our $\{C_{\rm n}\}$ values.

 \paragraph{}

\begin{figure}
\plotone{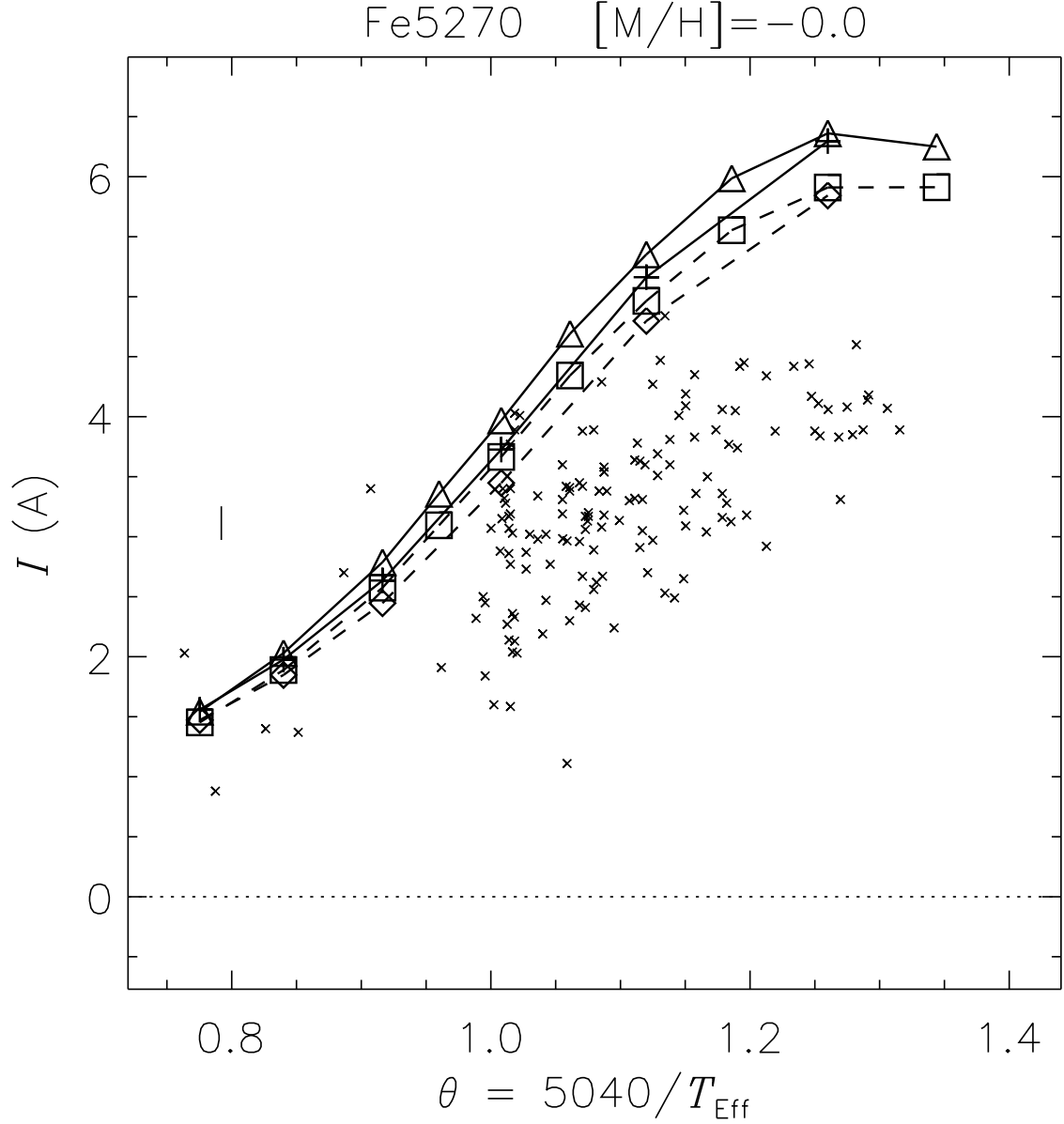}
\caption{Same as Fig. \ref{Fe5270_4000}, but for $I$ as a function of
$\theta = 5040/T_{\rm eff}$ at $[{{\rm M}\over {\rm H}}] = 0.0$.   
For comparison we show the observationally derived $I$ and 
$[{{\rm M}\over {\rm H}}]$ values of \citet{wortheyo97} for stars of $[{{\rm M}\over {\rm H}}]$ within $\pm 1.0$ 
(``X'' symbols). 
\label{Fe5270_0} }
\end{figure}

\begin{figure}
\plotone{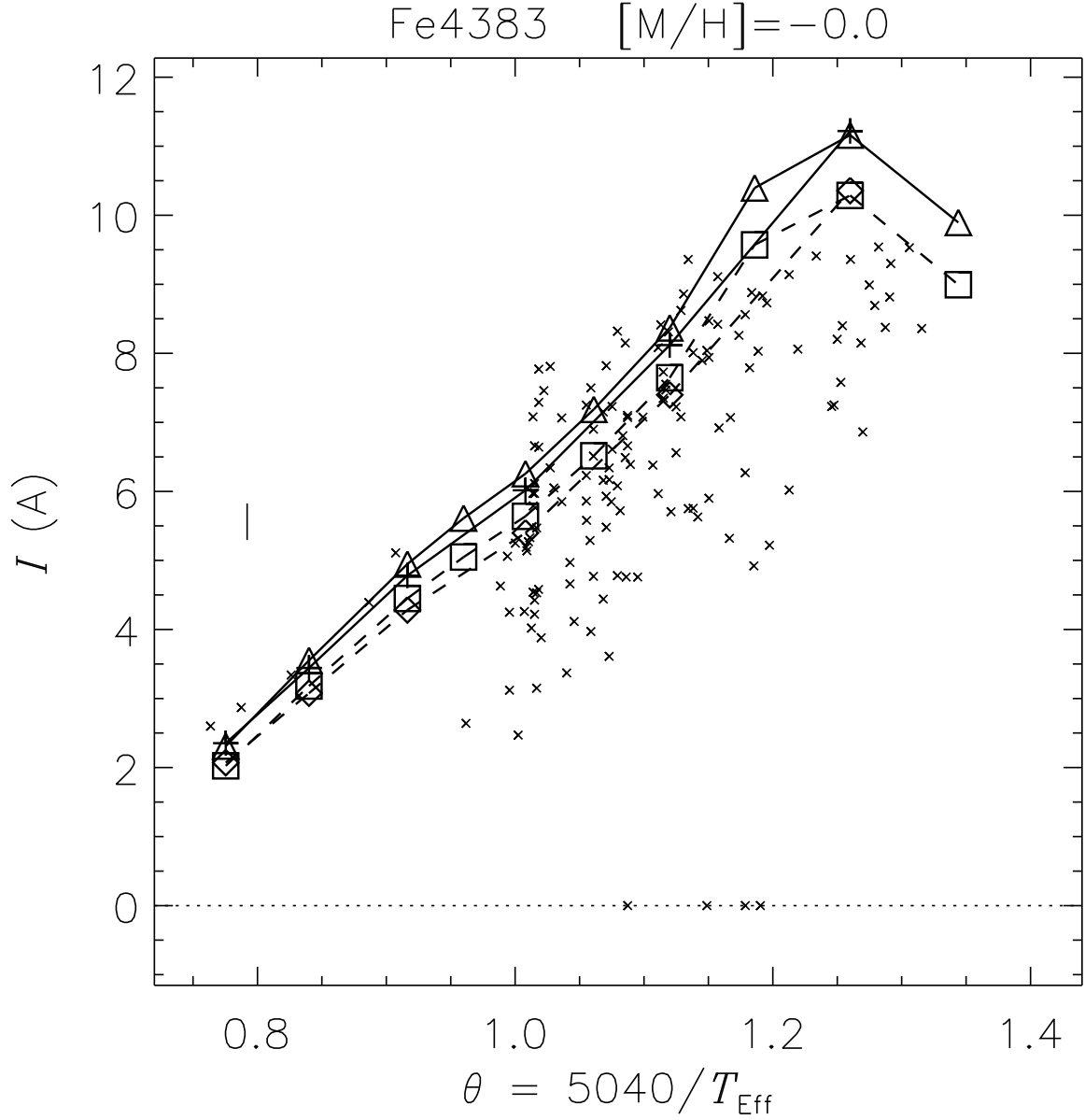}
\caption{Same as Fig. \ref{Fe5270_0}, but for index Fe4383.  $I$ is double valued at the cool 
edge of the grid, making its use there problematic, but is generally useful as an $[{{\rm M}\over {\rm H}}]$
diagnostic otherwise.
\label{Fe4383_0}}
\end{figure}

\begin{figure}
\plotone{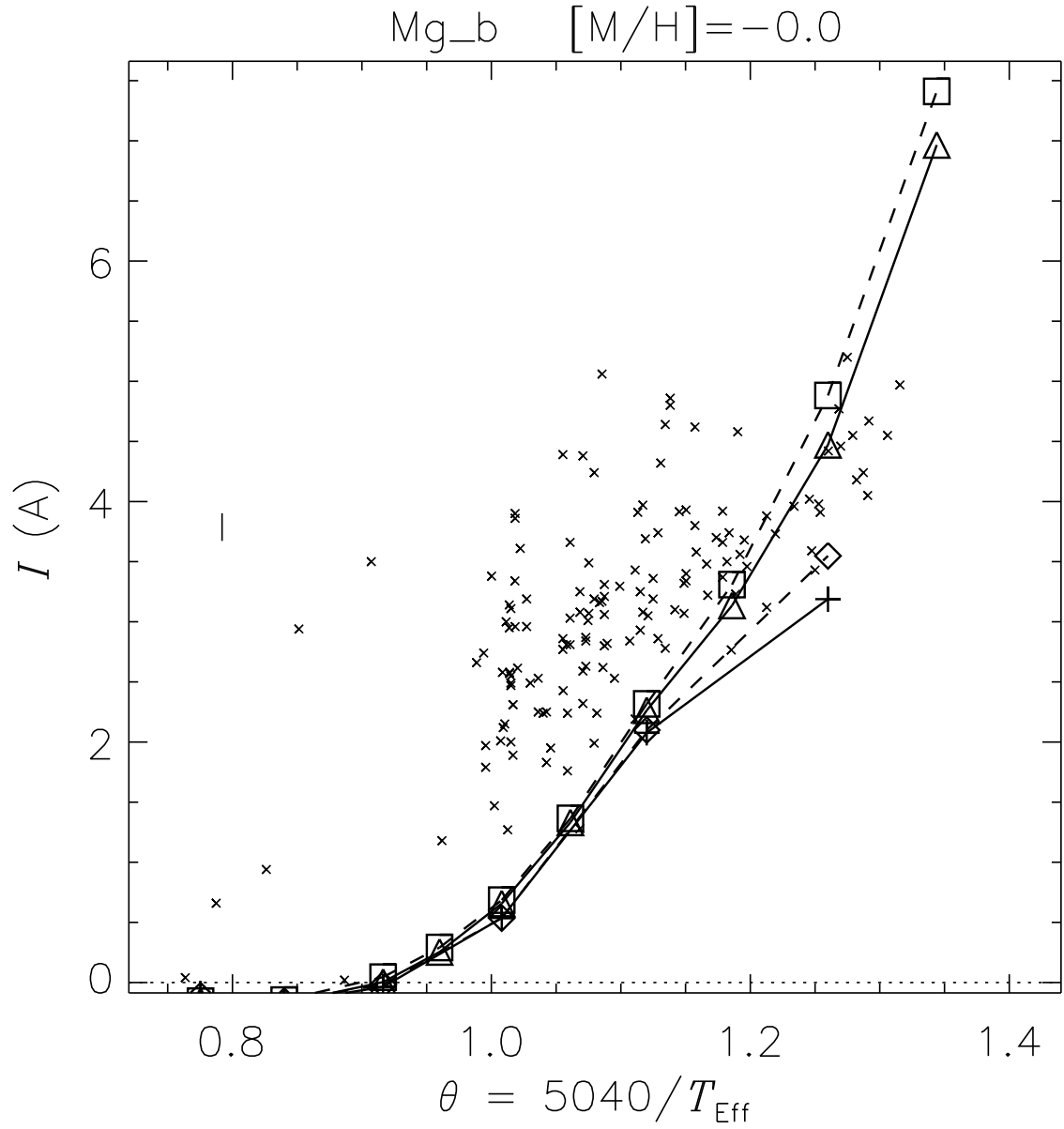}
\caption{Same as Fig. \ref{Fe5270_0}, but for index Mg $b$.
\label{Mgb_0}}
\end{figure}

\begin{figure}
\plotone{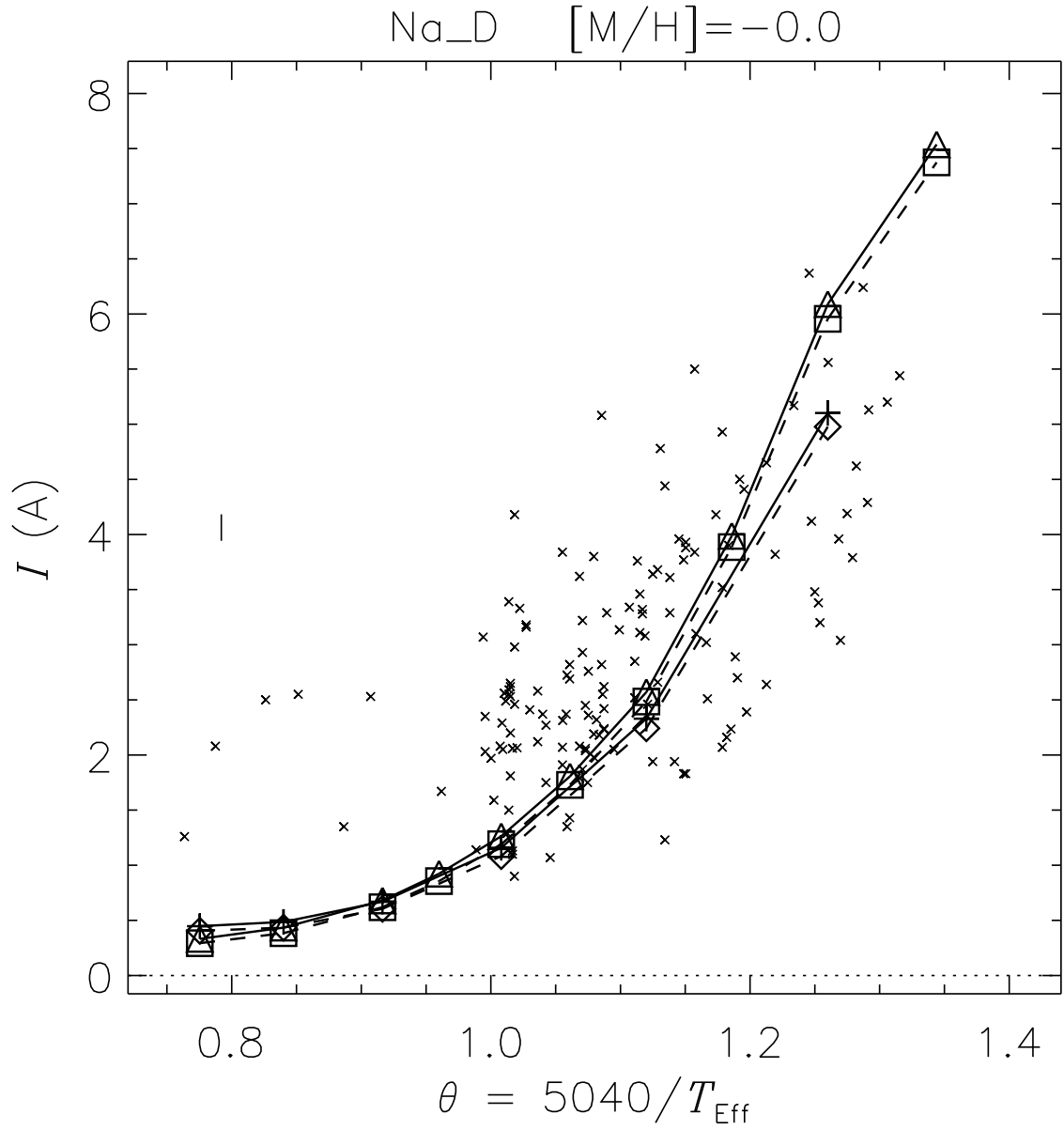}
\caption{Same as Fig. \ref{Fe5270_0}, but for index Na $D$.
\label{NaD_0}}
\end{figure}

\begin{figure}
\plotone{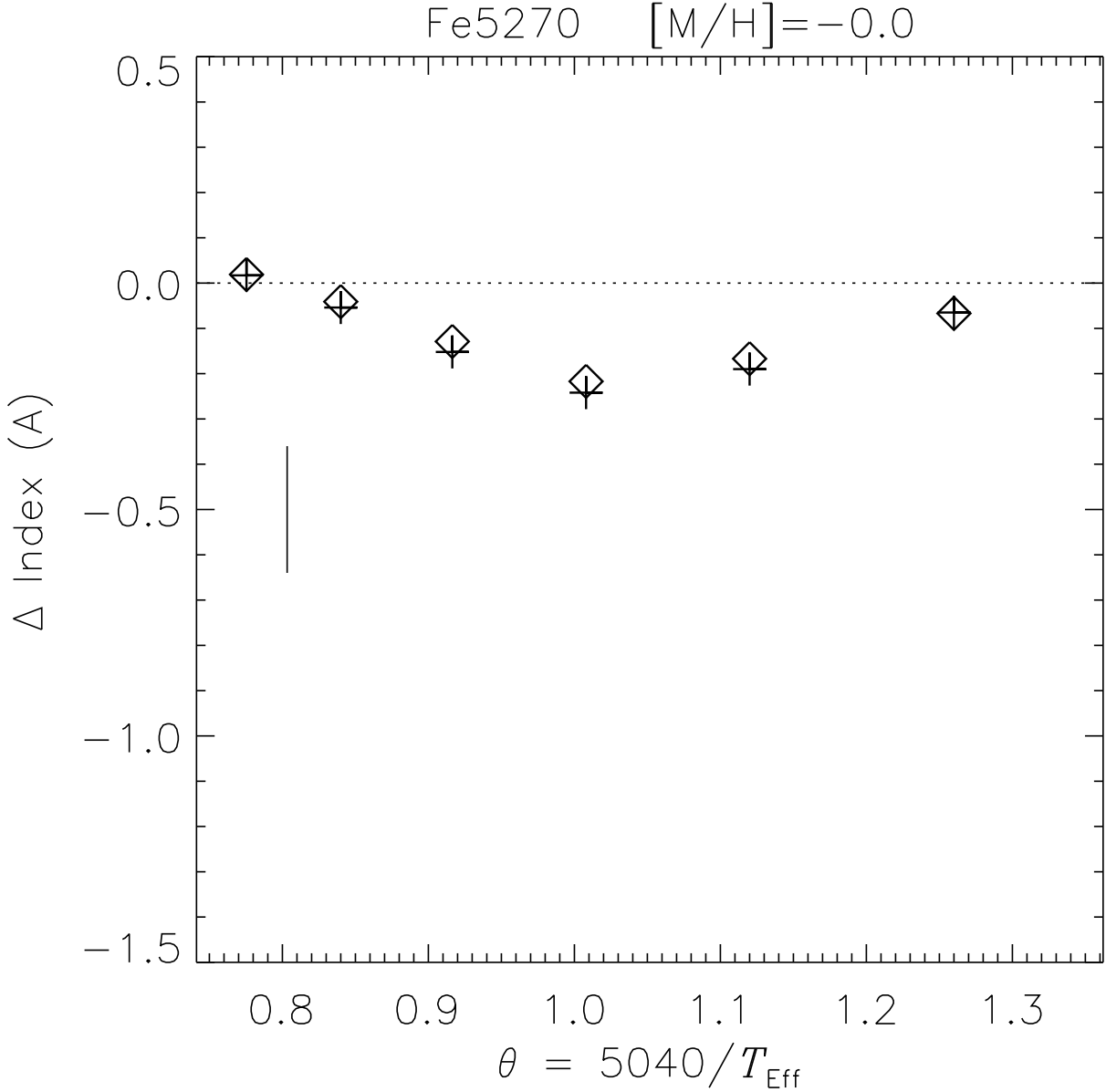}
\caption{Index Fe5270: The difference between $I$ computed in NLTE and that in LTE, $\Delta I = I_{\rm NLTE} - I_{\rm LTE}$,
as a function of $\theta = {5040\over T_{\rm eff}}$ for $[{{\rm M}\over {\rm H}}]=0.0$.  Results for $I$ derived from 
spectra of SDSS (crosses)
and IDS (diamonds) resolution, $R$.  The horizontal dotted line indicates a $\Delta I$
value of zero. 
\label{Fe5270diff}}
\end{figure}

\begin{figure}
\plotone{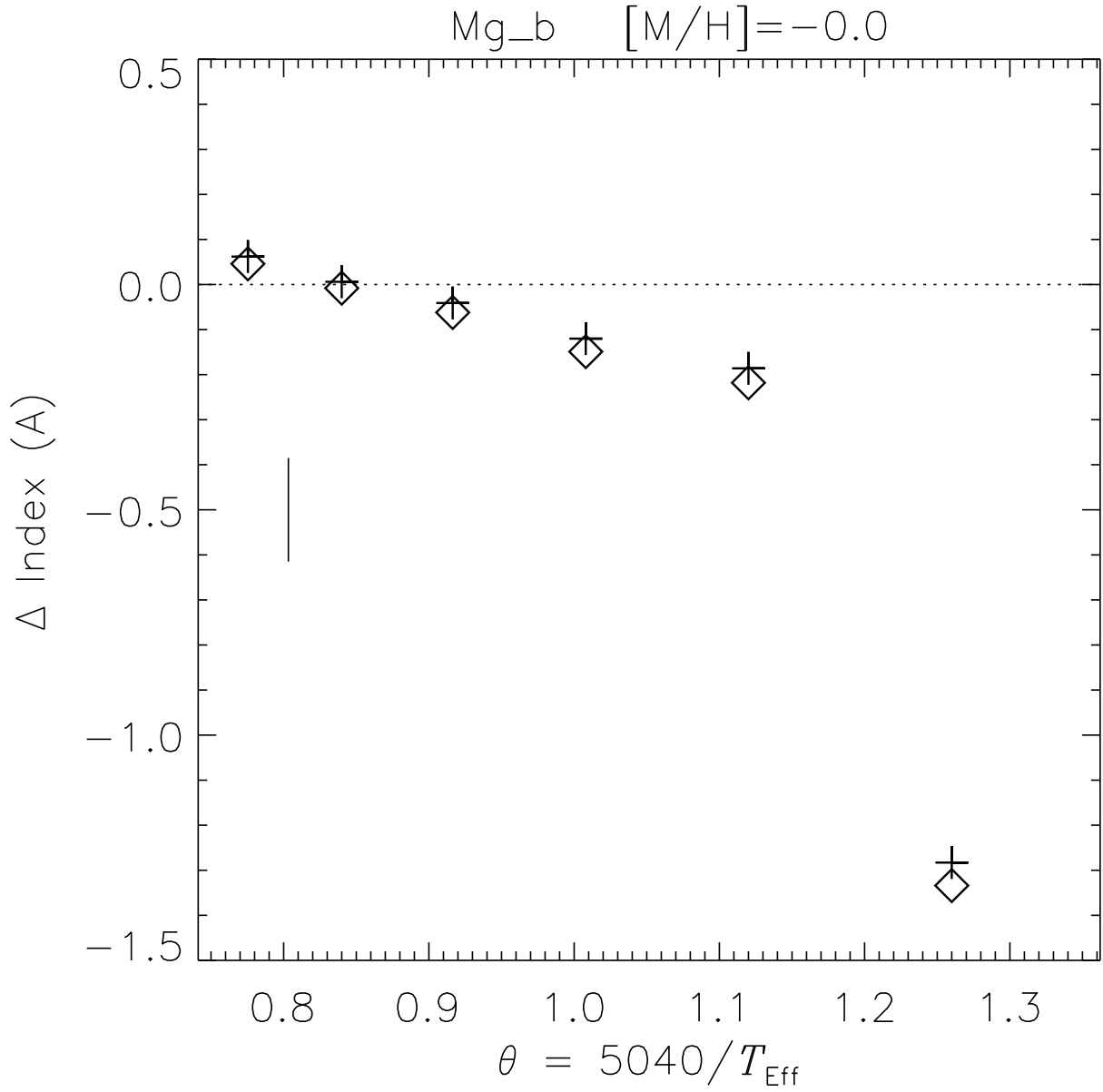}
\caption{Same as Fig. \ref{Fe5270diff}, except for index Mg $b$. \label{Mgbdiff}}
\end{figure}

\begin{figure}
\plotone{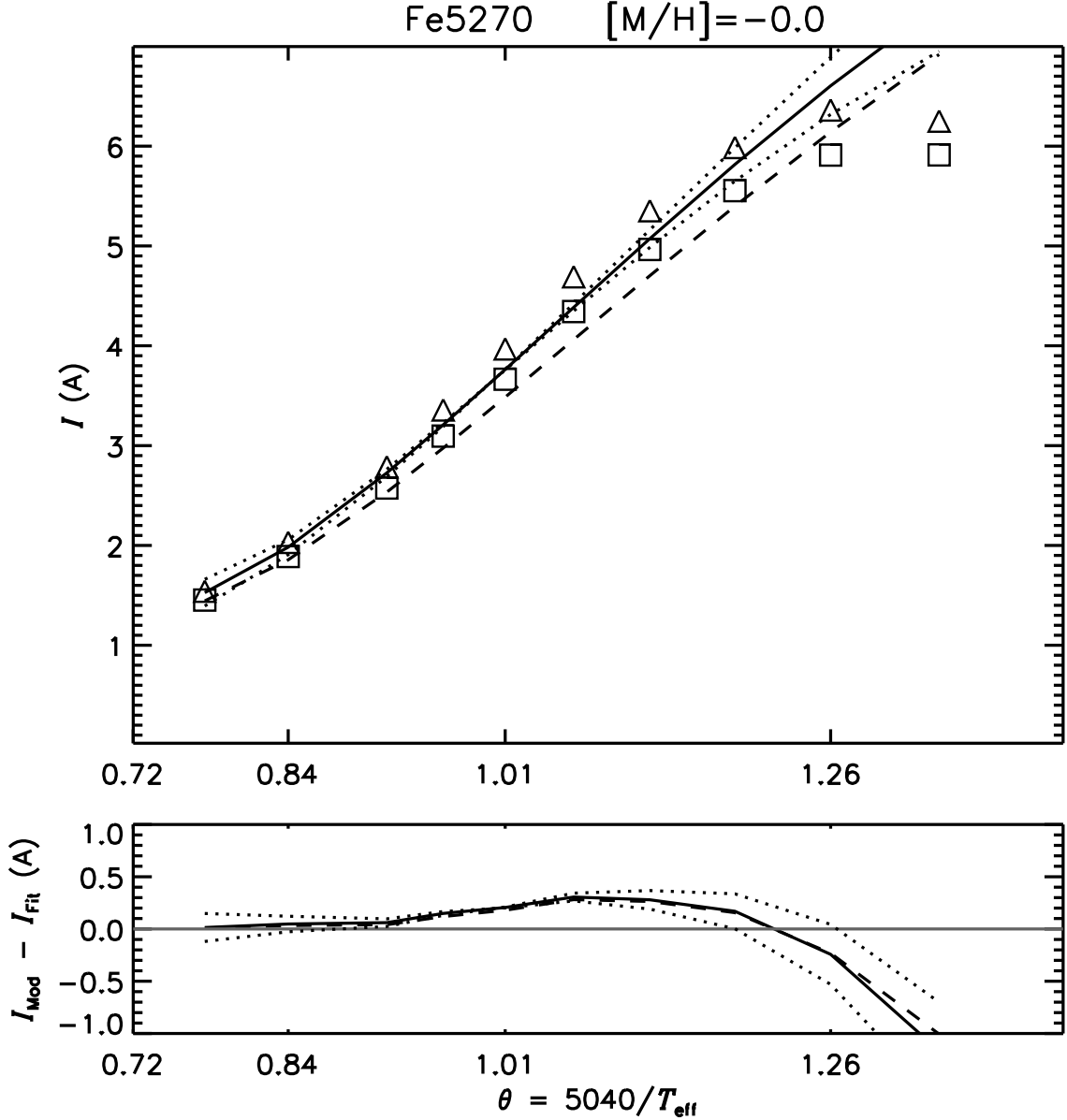}
\caption{Fe5270:  {\bf $I(\theta)$ based on} multiple linear regression fits for $I(\log\theta)$ at $[{{\rm M}\over {\rm H}}]=0.0$.  
Upper panel: Fitted relation (solid line) to modeled $I$ values (triangles) at SDSS resolution, and 
similarly at IDS resolution (dashed line and squares).  The dotted lines show the fitted relation
for SDSS resolution computed with fitting coefficients, $C_{\rm n}$, that have had their
1-$\sigma$ error from the $\chi^2$ fitting procedure (see text) added and subtracted from them. 
Lower panel: The residual
values for SDSS and IDS resolution (solid and dashed lines, respectively).  The dotted lines are
the residuals for SDSS resolution for a fitted relation computed with $C_{\rm n}$ values with 
their 1-$\sigma$ errors added and subtracted (see upper panel caption).  The horizontal gray 
line indicates a residual of zero. 
\label{residTeffFe5270}}
\end{figure}

\begin{figure}
\plotone{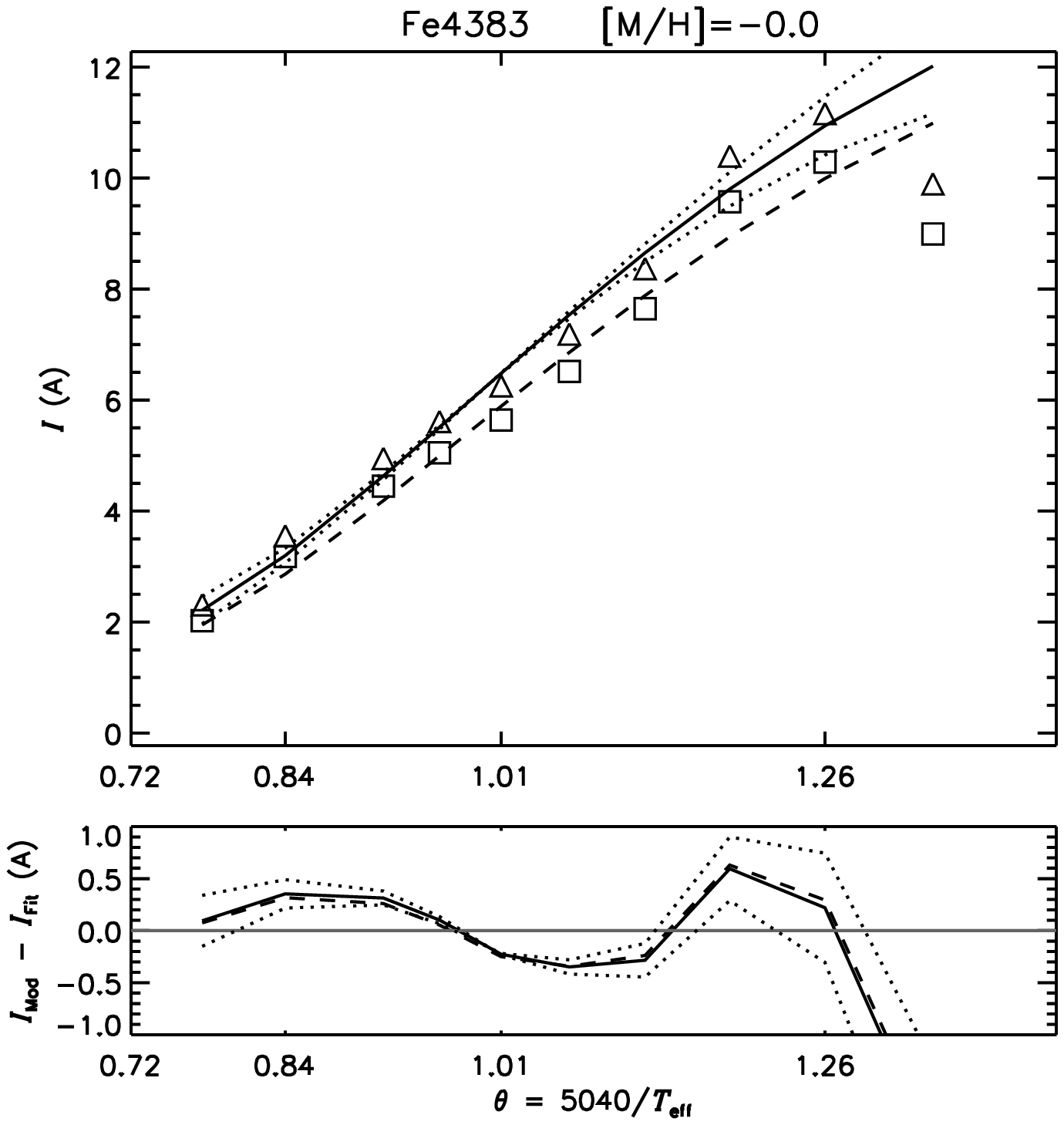}
\caption{Same as Fig. \ref{residTeffFe5270}, but for Fe4383.
\label{residTeffFe4383}}
\end{figure}

\begin{figure}
\plotone{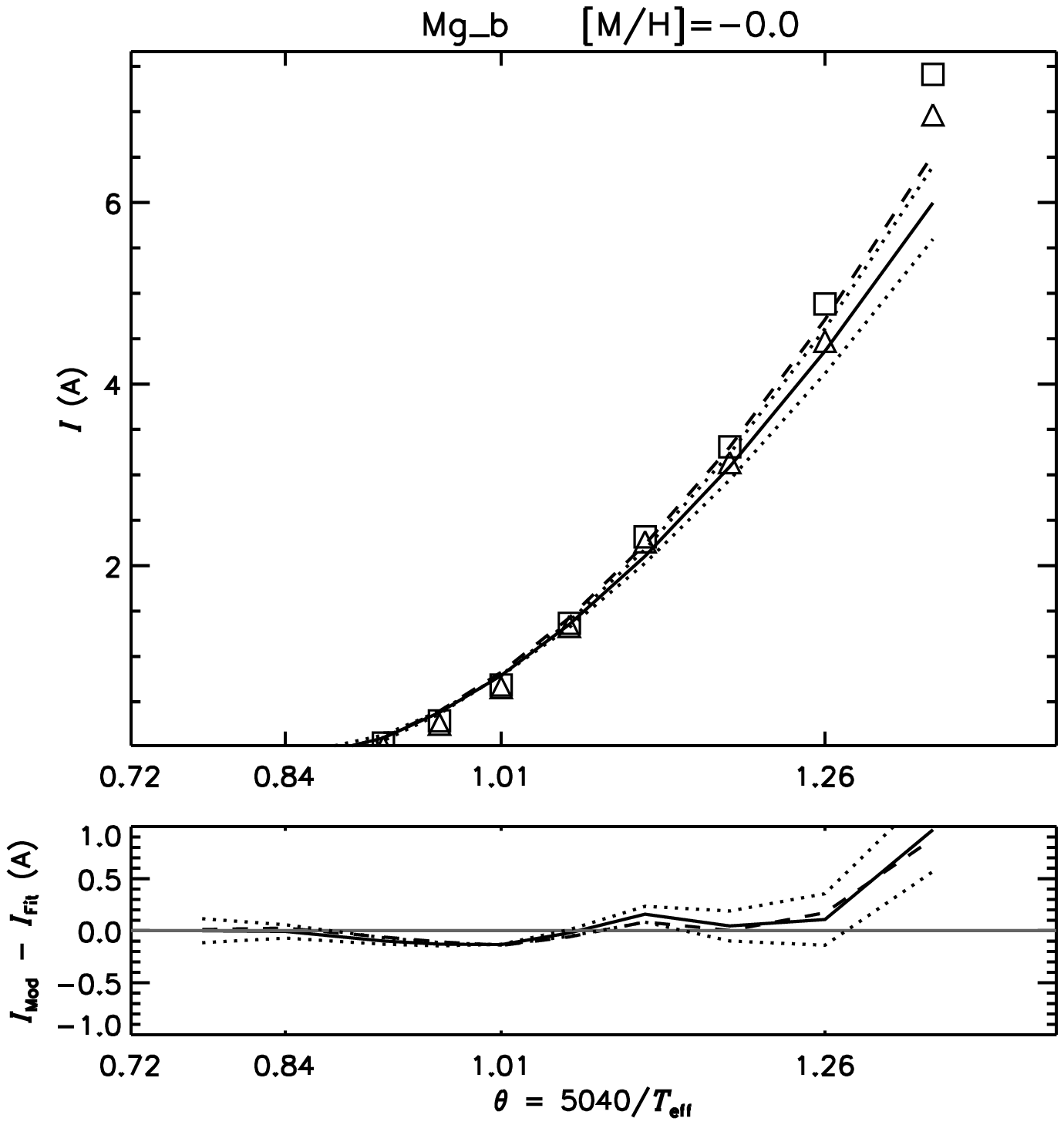}
\caption{Same as Fig. \ref{residTeffFe5270}, but for Mg $b$.
\label{residTeffMgb}}
\end{figure}

\begin{figure}
\plotone{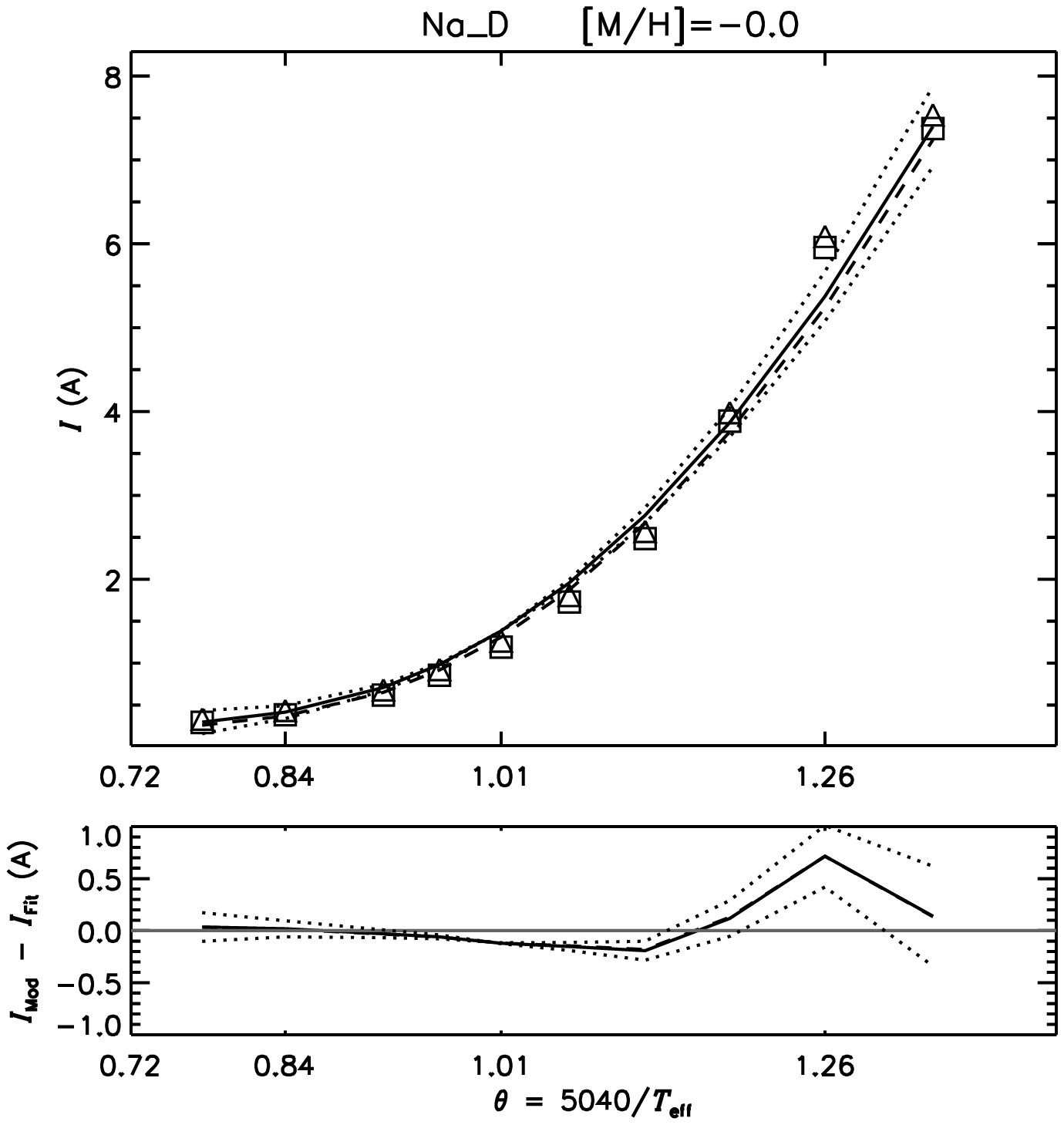}
\caption{Same as Fig. \ref{residTeffFe5270}, but for Na $D$.
\label{residTeffNaD}}
\end{figure}

\begin{figure}
\plotone{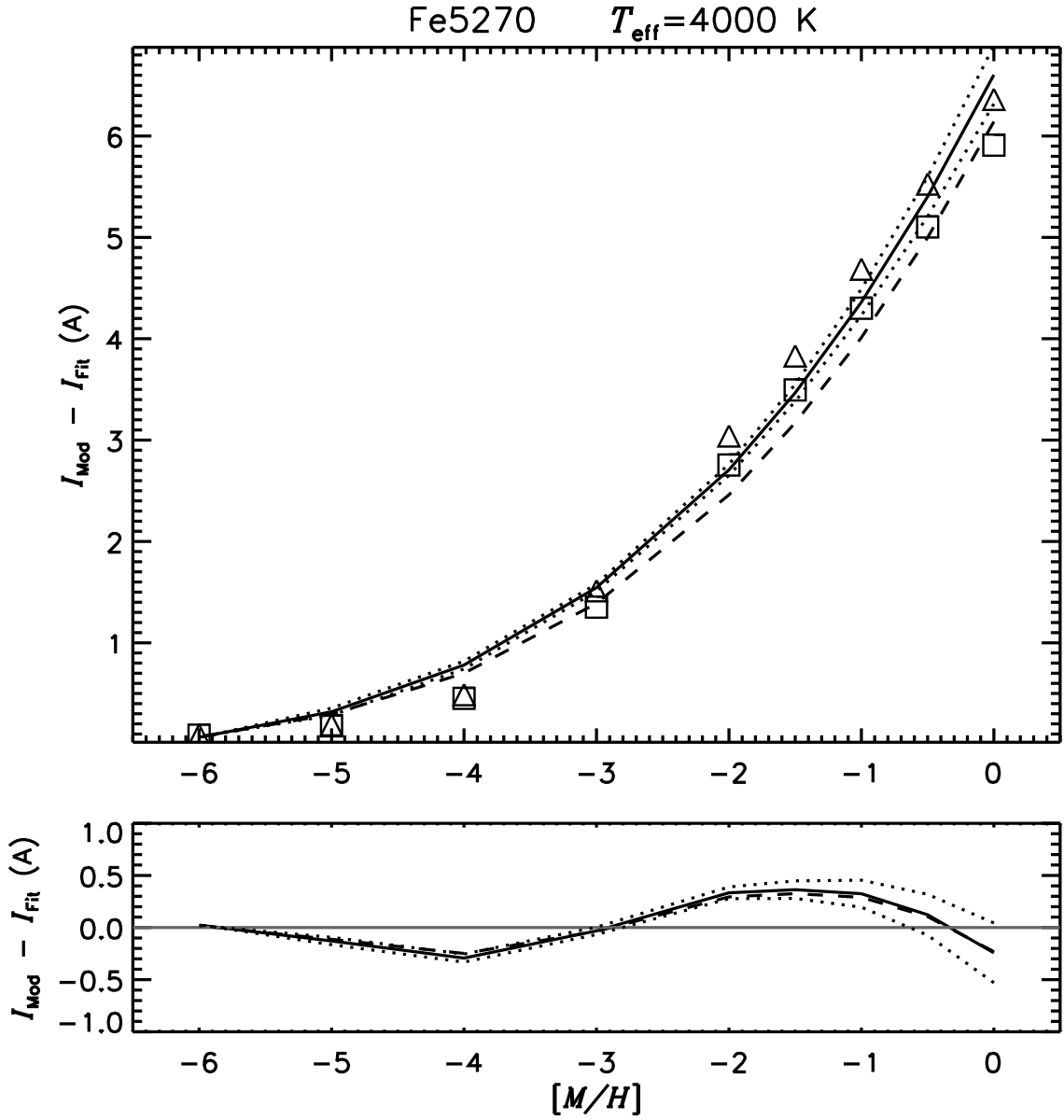}
\caption{Same as Fig. \ref{residTeffFe5270}, but for the $I([{{\rm M}\over {\rm H}}])$ 
relation at $T_{\rm eff}=4000$ K. 
\label{residAbndFe5270}}
\end{figure}

\begin{figure}
\plotone{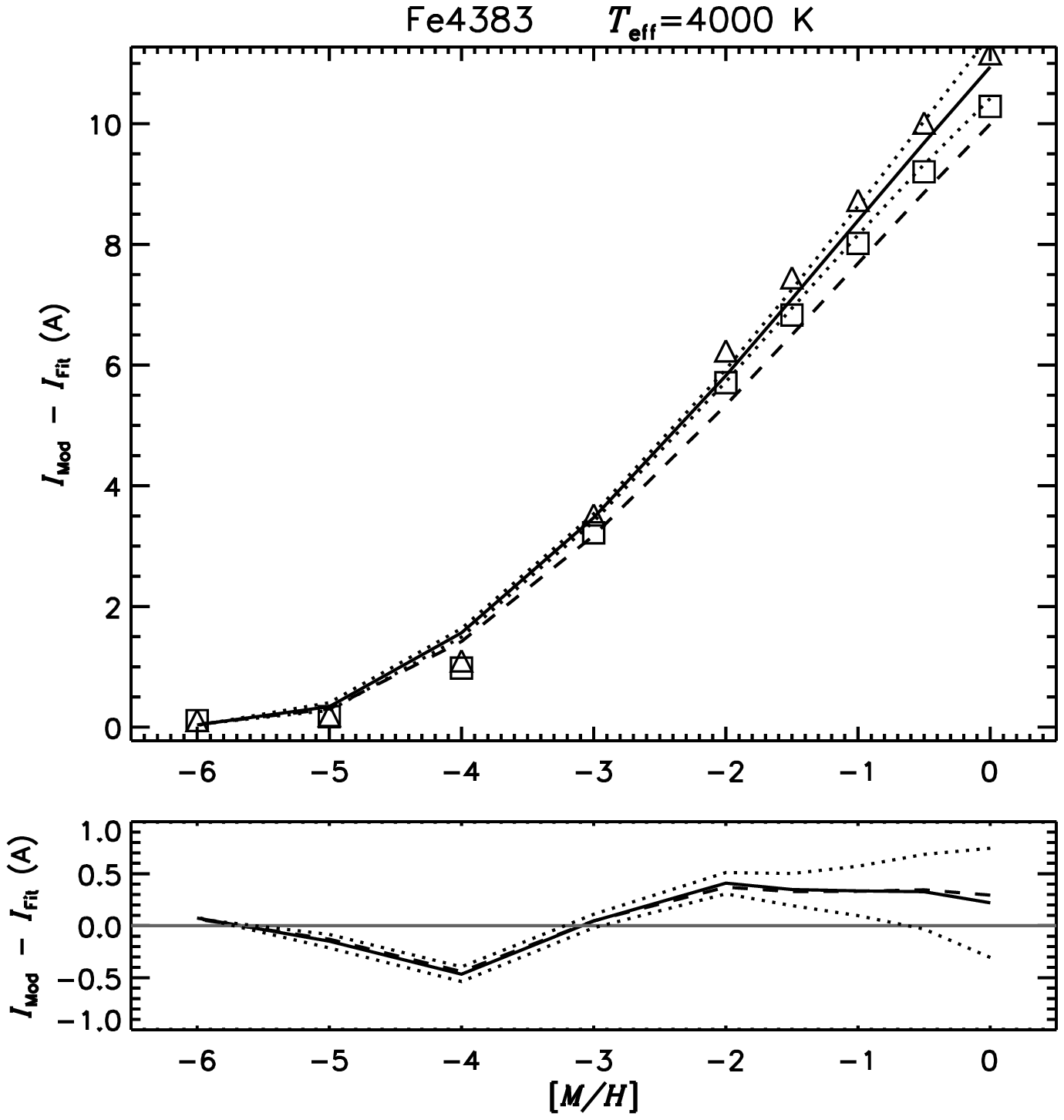}
\caption{Same as Fig. \ref{residAbndFe5270}, but for Fe4383.
\label{residAbndFe4383}}
\end{figure}

\begin{figure}
\plotone{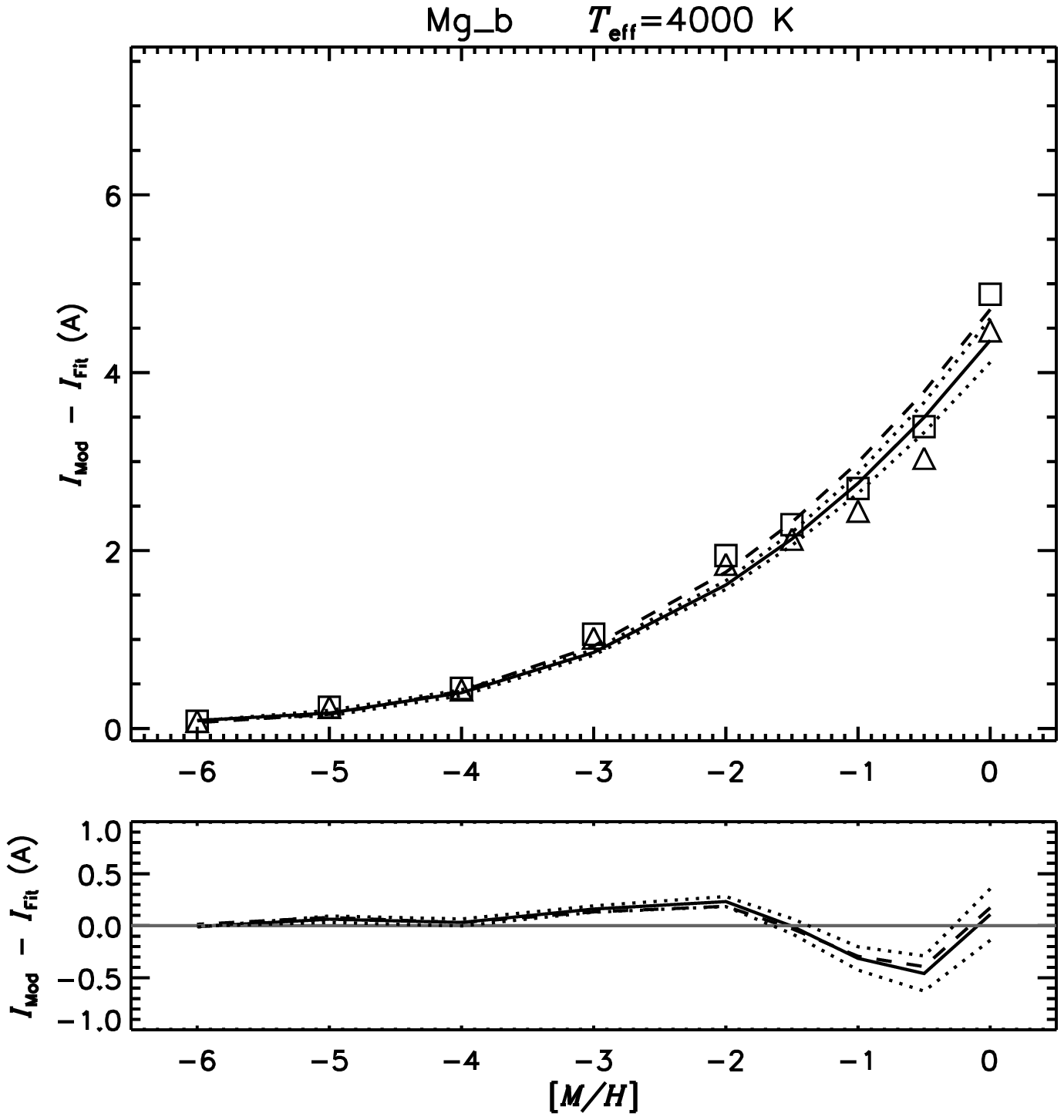}
\caption{Same as Fig. \ref{residAbndFe5270}, but for Mg $b$.
\label{residAbndMgb}}
\end{figure}

\begin{figure}
\plotone{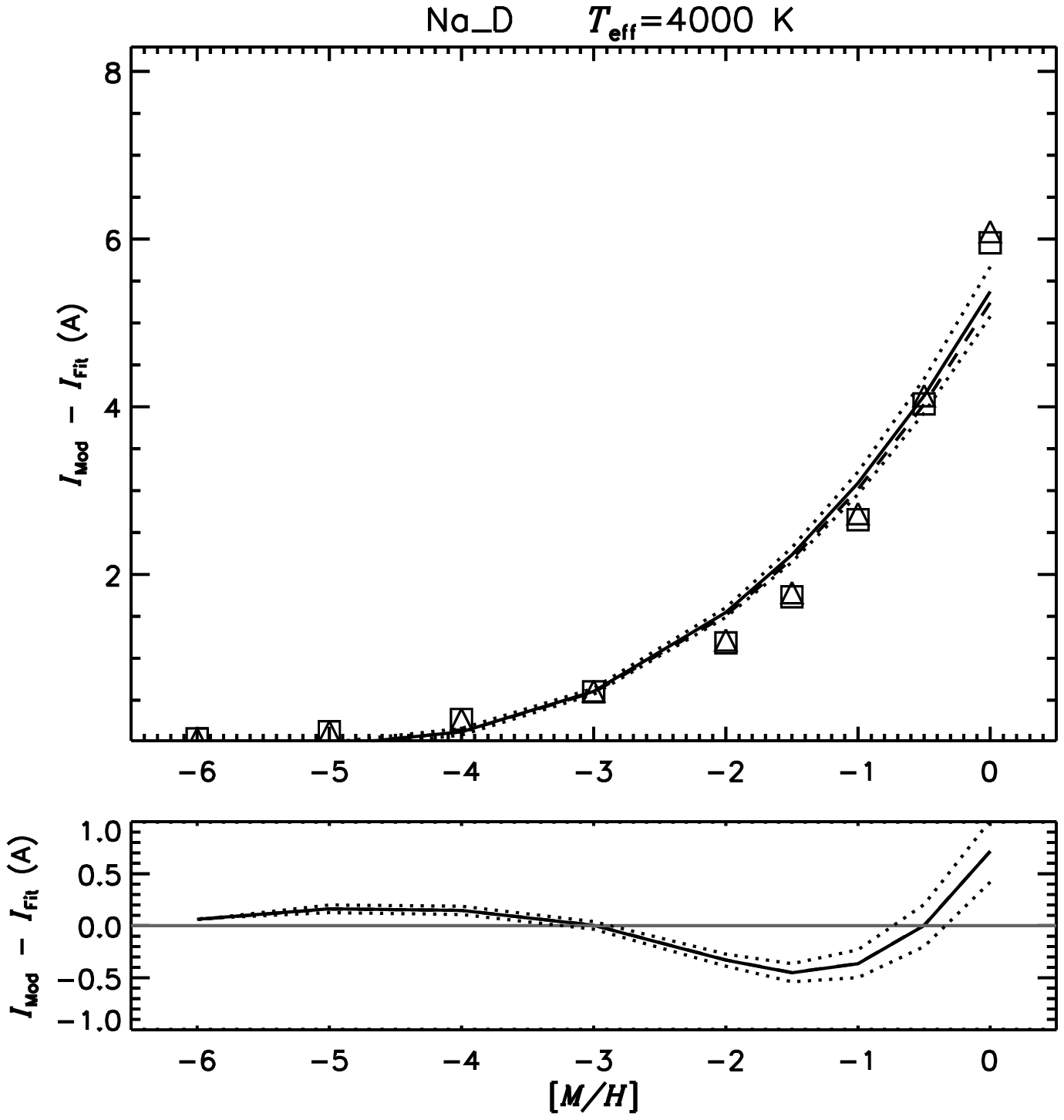}
\caption{Same as Fig. \ref{residAbndFe5270}, but for Na $D$.
\label{residAbndNaD}}
\end{figure}

  For Fe-dominated Lick indices, the effect of NLTE is to generally weaken the value of $I$ 
at any give ($T_{\rm eff}, [{{\rm M}\over {\rm H}}]$) value.  To put the magnitude of the
NLTE effect into context, the change, $\Delta I$,
caused by NLTE effects is generally comparable to the change that results from
computing $I$ from spectra of SDSS resolution rather than that of IDS.  The partial
derivatives can be used to estimate a change in inferred $T_{\rm eff}$ at fixed $[{{\rm M}\over {\rm H}}]$,
or a change in inferred $[{{\rm M}\over {\rm H}}]$ at fixed $T_{\rm eff}$, resulting from 
a change in $I$ ({\it e.g.} as caused by NLTE effects) at any ($T_{\rm eff}, [{{\rm M}\over {\rm H}}]$)
value pair throughout our grid.  For example, from Figs. \ref{Fe5270_0} and \ref{Fe4383_0}, 
for stars of inferred $T_{\rm eff} \gtrsim 4200$ K ($\theta \lesssim 1.2$), an Fe-dominated $I$ value 
computed in LTE that is too strong might be compensated for by inferring a $T_{\rm eff}$ value that
is too large, for fixed inferred $[{{\rm M}\over {\rm H}}]$.  



\acknowledgments

CIS is grateful for NSERC Discovery Program grant RGPIN-2014-03979.  The calculations were
performed with the facilities of the Atlantic Computational Excellence Network (ACEnet).

\clearpage



\clearpage

\clearpage

\clearpage

\clearpage

\clearpage

\clearpage

\clearpage

\clearpage

\clearpage

\clearpage

\clearpage

\clearpage

\clearpage

\clearpage

\clearpage

\clearpage

\clearpage

\clearpage

\clearpage

\clearpage

\clearpage

\clearpage

\begin{deluxetable}{lrrrrrrr}
\tablecolumns{8}
\tablecaption{Index values, $I$, at SDSS spectral resolution from LTE models for the nine Lick-XMP indices (see text) for models 
of $T_{\rm eff}=4000$ K and 
$\log g=2.0$, in which the $I$ values are especially large, and select $[{{\rm Fe}\over {\rm H}}]$ values computed with
scaled solar abundances (first number in each data pair) and with a relative enhancement of $\alpha$-element abundances 
by +0.4 (second number in each data pair).  We note that for the scaled-solar abundance models,  $[{{\rm Fe}\over {\rm H}}]$
is identical to our $[{{\rm M}\over {\rm H}}]$ grid parameter.} 
\tablehead{
\colhead{ } & \multicolumn{7}{c}{Lick index, $I$}                 \\ 
}
\startdata
$[{{\rm Fe}\over {\rm H}}]$  &       Fe4383  &      Fe4531  &     Fe5015  &     Mg$_{\rm 1}$ &   Mg $b$     \\
-2.0                         & 6.235, 5.454  & 4.041, 4.302 & 5.167, 5.257 & 0.189, 0.275    & 1.845, 2.449    \\
-4.0                         & 1.098, 1.002  & 0.520, 0.716 & 0.647, 0.891 & 0.006, 0.016    & 0.433, 0.670    \\
   \\
$[{{\rm Fe}\over {\rm H}}]$  &       Fe5270 &      Fe5335  &      Fe5406 &         Na $D$    \\
-2.0                         & 3.040, 2.988 & 2.635, 2.523 & 1.923, 1.797 & 1.217, 1.054 \\       
-4.0                         & 0.488, 0.545 & 0.500, 0.546 & 0.374, 0.404 & 0.271, 0.288 \\
\hline\\
\enddata
\label{alphaTab}
\end{deluxetable}

\clearpage

\begin{deluxetable}{lrrrr}
\tablecolumns{5}
\tablecaption{Range of $[{{\rm M}\over{\rm H}}]$ value for which Lick index, $I$, is a significant $[{{\rm M}\over{\rm H}}]$ 
diagnostic for differences of $\Delta [{{\rm M}\over{\rm H}}]\approx 1$ for select $T_{\rm eff}$ values, and corresponding
 $(V-K)$ values at $[{{\rm M}\over{\rm H}}]=0.0$, at SDSS spectral resolution. }
\tablehead{
\colhead{} & \multicolumn{4}{c}{$T_{\rm eff}$ (K)}                 \\ 
\colhead{} & \colhead{3750} & \colhead{4500} & \colhead{5000} & \colhead{6500}   \\ 
\colhead{} & \multicolumn{4}{c}{$V-K$ at $[{{\rm M}\over{\rm H}}]=0.0$}                         \\ 
\colhead{Index} & \colhead{4.33} & \colhead{2.58} & \colhead{2.02} & \colhead{0.89}   \\ 
} 
\startdata
CN$_{\rm 2}$            & -5.0, -2.0 & -1.5, 0.0  & -0.5, 0.0  & is        \\  
Ca4227                  & -4.0, 0.0  & mv, neg    & mv, neg    & -3.0, 0.0 \\ 
G4300                   &  mv        & -3.0, -5.0 & -1.5, -4.0 & 0.0, -3.0 \\   
Fe4383\tablenotemark{*}        & -5.0, -1.0 & -4.0, 0.0  & -4.0, -1.0 & -3.0, 0.0 \\
Ca4455                  & -4.0, 0.0  & -3.0, 0.0  & -3.0, 0.0  & -2.0, 0.0 \\  
Fe4531\tablenotemark{*}        & -5.0, 0.0  & -4.0, 0.0  & -3.0, 0.0  & -3.0, 0.0 \\
Fe4668                  & -1.0, 0.0  & -1.5, 0.0  & -1.5, 0.0  & -1.0, 0.0 \\
Fe5015\tablenotemark{*}        & -5.0, 0.0  & -4.0, 0.0  & -4.0, 0.0  & -3.0, 0.0 \\
Mg$_{\rm 1}$\tablenotemark{*}  & -5.0, -2.0 & -3.0, 0.0  & -2.0, 0.0  & -1.0, 0.0 \\   
Mg $b$\tablenotemark{*}           & -6.0, 0.0  & -5.0, 0.0  & mv      & mv \\ 
Fe5270\tablenotemark{*}        & -6.0, 0.0  & -4.0, 0.0  & -3.0, 0.0  & -2.0, 0.0 \\ 
Fe5335\tablenotemark{*}        & -6.0, 0.0  & -5.0, 0.0  & -4.0, 0.0  & -3.0, 0.0 \\
Fe5406\tablenotemark{*}        & -6.0, 0.0  & -5.0, 0.0  & -3.0, 0.0  & -2.0, 0.0 \\
Fe5709                  & -4.0, 0.0  & -3.0, 0.0  & -3.0, 0.0  & -1.0, 0.0 \\
Fe5782                  & -3.0, -1.0 & -2.0, 0.0  & -2.0, 0.0  & -1.0, 0.0 \\ 
Na $D$\tablenotemark{*}           & -6.0, 0.0  & -5.0, 0.0  & -4.0, 0.0  & is \\ 
TiO$_{\rm 1}$           & -3.0, 0.0  & -1.0, 0.0  & -1.0, 0.0  & is \\
TiO$_{\rm 2}$           & -4.0, 0.0  & -3.0, 0.0  & -2.0, 0.0  & -1.0, 0.0 \\
\hline\\
\enddata
\tablenotetext{*}{$I$ is a strong, well-behaved $[{{\rm M}\over{\rm H}}]$-indicator over a broad $[{{\rm M}\over{\rm H}}]$ range down to XMP values (at least $[{{\rm M}\over{\rm H}}] \le -5.0$) for at least some GK star $T_{\rm eff}$ values.} 
is, mv, neg: $I$ suffers from one or more of several pathologies over a significant range of $[{{\rm M}\over{\rm H}}]$: 
is: {$I$ is insensitive to $[{{\rm M}\over{\rm H}}]$ as judged by the corresponding value of $\sigma_{\rm Worthey}$ (see text).}
mv: {$I$ is multi-valued.}
neg: {$I$ is negative (for those that are in linear $W_\lambda$ units).}
\label{XMPIs}
\end{deluxetable}

\clearpage

\begin{deluxetable}{lrrrrrrr}
\tablecolumns{8}
\tablecaption{LTE fitting function coefficients, $C_{\rm n}$, for independent variables $\log\theta$ and $[{{\rm M}\over {\rm H}}]$ for stars of 
$3750 < T_{\rm eff} < 6500$ K at IDS spectral resolution for our nine Lick-XMP indices.  Quantities in brackets below each fitted
$C_{\rm n}$ value are the 1-$\sigma$ ``fitting errors'' (see text) estimated from $\chi^2$, and may be interpreted as $68\%$ confidence
intervals.}
\tablehead{
\colhead{Polynomial term} & \multicolumn{7}{c}{Lick index, $I$}                 \\ 
} 
\startdata
           Term ($n$)               &        Fe4383  &      Fe4531  &     Fe5015  &      Mg$_{\rm 1}$&   Mg $b$     \\
    $C_{\rm 0}$                     &        5.7407   &      5.1054  &    8.5322  &      0.0547      &   0.7475    \\
   $\log\theta$                     &       42.7841  &     26.6568  &    38.7751  &      1.6149      &  22.3066    \\
    $\sigma$                        &       (1.9465)   &     (0.9536)   &    (1.1653)   &      (0.1791)      &  (0.9649)    \\
 $\log^2 \theta$                     &       34.3084  &     47.8340  &    32.7787  &     16.8150      & 150.1230   \\
    $\sigma$                        &       (14.3478)   &     (7.0287)   &    (8.5895)   &     (1.3200)       &  (7.1125)   \\ 
 $\log^3 \theta$                     &     -383.6426  &      6.9564  &   -16.5491  &     54.8095      & 208.7209   \\
    $\sigma$                        &       (135.1847)  &      (66.2243) &   (80.9304)   &     (12.4368)      & (67.0145)   \\  
$[{{\rm M}\over{\rm H}}]$           &        1.3678  &      2.2633  &     4.9326  &      0.0345      &   0.5357   \\
    $\sigma$                        &       (0.1858)   &      (0.0910)  &    (0.1112)   &     (0.0171)       & (0.0921)    \\ 
$[{{\rm M}\over{\rm H}}]^2$         &       -0.1555  &      0.2680  &     0.9472  &      0.0021      &   0.1526     \\
    $\sigma$                        &        (0.0749)   &      (0.0367)  &    (0.0448)   &      (0.0069)      &  (0.0371)   \\ 
$[{{\rm M}\over{\rm H}}]^3$         &       -0.0376  &      0.0050  &     0.0600  &     -0.0005      &   0.0132    \\
    $\sigma$                        &        (0.0083)   &      (0.0041)  &    (0.0050)   &     (0.0008)       &  (0.0041)   \\    
$[{{\rm M}\over{\rm H}}]\log\theta$ &       8.1137  &      7.0252  &    13.5876  &      0.7520      &  11.3125    \\
    $\sigma$                        &        (1.1698)  &      (0.5731)  &    (0.7003)   &      (0.1076)      &    (0.5799)  \\ 
$[{{\rm M}\over{\rm H}}]^2\log\theta$ &      0.2170  &      0.3786  &     1.1877  &      0.0629      &   1.2825   \\
    $\sigma$                        &        (0.1900)   &   (0.0931)     &    (0.1138)   &     (0.0175)      &   (0.0942)    \\  
$[{{\rm M}\over{\rm H}}]\log^2 \theta$ &     2.8481  &      7.2659  &     5.3031  &      3.4235      &  31.2653    \\
    $\sigma$                        &        (4.3128)   &      (2.1128)  &    (2.5820)   &     (0.3968)      &   (2.1380)   \\ 
Total $\sigma$                       &       136.0316  &    66.6392  &      81.4374  &   12.5147      &  67.4343   \\
Red $\chi^2$                           &        0.1755 &        0.0421  &     0.0629  &     0.0015    &    0.0431\\
   \\
           Term ($n$)              &          Fe5270 &      Fe5335  &      Fe5406 &         Na $D$   \\
    $C_{\rm 0}$                    &         3.3993  &     3.5991   &     1.9507  &      1.2374    \\
   $\log\theta$                    &        24.7325  &    22.7161   &    19.5786  &     20.5914   \\
    $\sigma$                        &        (1.0332)  &     (0.7873)   &     (0.5449)  &     (1.2010)  \\ 
 $\log^2 \theta$                    &        43.8640  &   27.8599   &    55.6521  &    151.3475   \\
    $\sigma$                        &        (7.6155)   &   (5.8036)    &    (4.0165)   &     (8.8525)  \\ 
 $\log^3 \theta$                    &      -179.7178  &  -169.1160   &   -68.7232  &    407.8087   \\
    $\sigma$                        &       (71.7529)   &  (54.6813)     &   (37.8434)   &      (83.4082)  \\ 
$[{{\rm M}\over{\rm H}}]$          &         1.7790  &     1.8678   &     1.1233  &      1.0780    \\
    $\sigma$                        &        (0.0986)   & (0.0752)       &    (0.0520)   &      (0.1146)  \\ 
$[{{\rm M}\over{\rm H}}]^2$        &         0.3000  &     0.3264   &     0.2176  &      0.2884    \\
    $\sigma$                        &        (0.0398)   &  (0.0303)      &     (0.0210)  &      (0.0462)   \\ 
$[{{\rm M}\over{\rm H}}]^3$        &         0.0165  &     0.0193   &     0.0140  &      0.0228    \\
    $\sigma$                        &        (0.0044)   &  (0.0033)     &   (0.0023)     &     (0.0051)   \\ 
$[{{\rm M}\over{\rm H}}]\log\theta$   &     6.3203  &     6.8152   &     6.9676  &     12.0989   \\
    $\sigma$                        &        (0.6209)   &    (0.4732)   &     (0.3275)   &      (0.7218)  \\ 
$[{{\rm M}\over{\rm H}}]^2\log\theta$ &      0.3882  &     0.5507   &     0.6483  &      1.4020    \\
    $\sigma$                        &        (0.1009)   &   (0.0769)     &   (0.0532)    &       (0.1172)  \\ 
$[{{\rm M}\over{\rm H}}]\log^2 \theta$ &     3.9294  &     1.8886   &    8.9946  &     33.6358   \\
    $\sigma$                        &         (2.2892)  &    (1.7445)    &    (1.2073)   &    (2.6610)  \\ 
Total $\sigma$                         &     72.2025  &   55.0239    &    38.0804  &    83.9307   \\
Red $\chi^2$                           &        0.0494   &   0.0287   &    0.0137    &      0.0668   \\
      \\
\hline\\
\enddata
\label{coeffsthetaIDS}
\end{deluxetable}

\clearpage

\begin{deluxetable}{lrrrrrrr}
\tablecolumns{8}
\tablecaption{Same as Table \ref{coeffsthetaIDS}, but for spectra of SDSS spectral resolution.}
\tablehead{
\colhead{Polynomial term} & \multicolumn{7}{c}{Lick index, $I$}                 \\ 
} 
\startdata
           Term ($n$)               &        Fe4383  &      Fe4531  &     Fe5015  &      Mg$_{\rm 1}$&   Mg $b$     \\
    $C_{\rm 0}$                     &      6.3207    &    5.8713    &    9.9981   &     0.0596    &    0.7159     \\
   $\log\theta$                     &      46.5420   &    29.7759   &    47.4312  &      1.6445   &    21.1327    \\
    $\sigma$                        &      (2.1429)    &  (1.1107)    &  (1.0168)   &   (0.1790)    &  (1.0098)    \\
 $\log^2 \theta$                     &    37.7593    &   44.8405    &   37.9836   &    16.8348    &  136.5346    \\
    $\sigma$                        &     (15.7958)    &  (8.1873)    &  (7.4950)   &   (1.3194)    &  (7.4430)    \\
 $\log^3 \theta$                     &    -426.9500  &    -45.0185  &   -124.2286 &      54.8555  &    149.0733     \\
    $\sigma$                        &     (148.8280)   &  (77.1404)   &  (70.6182)  &   (12.4315)   &  (70.1279)    \\
$[{{\rm M}\over{\rm H}}]$           &     1.5700    &    2.5478    &    5.5798   &     0.0376    &    0.5076      \\
    $\sigma$                        &       (0.2046)   &   (0.1060)   &   (0.0971)  &    (0.0171)   &   (0.0964)     \\
$[{{\rm M}\over{\rm H}}]^2$         &     -0.1475   &     0.2888   &     1.0275  &      0.0028   &     0.1489    \\
    $\sigma$                        &     (0.0825)     & (0.0427)     & (0.0391)    &   (0.0069)     &  (0.0389)    \\
$[{{\rm M}\over{\rm H}}]^3$         &    -0.0392    &    0.0041    &    0.0624   &    -0.0004    &    0.0134    \\
    $\sigma$                        &      (0.0091)    &  (0.0047)    &  (0.0043)   &    (0.0008)    &  (0.0043)     \\
$[{{\rm M}\over{\rm H}}]\log\theta$ &      8.5967   &     7.4703   &    15.1617  &      0.7638   &    10.7092     \\
    $\sigma$                        &      (1.2879)    &  (0.6675)    &  (0.6111)   &    (0.1076)    &  (0.6068)    \\
$[{{\rm M}\over{\rm H}}]^2\log\theta$ &   0.1880    &    0.3720    &    1.2213   &     0.0641    &    1.2324    \\
    $\sigma$                        &      (0.2092)    &  (0.1084)    &  (0.0993)   &    (0.0175)    &  (0.0986)    \\
$[{{\rm M}\over{\rm H}}]\log^2 \theta$ &  2.3983    &    5.9506    &    4.2027   &     3.4272    &   28.2502       \\
    $\sigma$                        &      (4.7481)    &  (2.4610)    &  (2.2530)   &   (0.3966)    &  (2.2373)      \\
Total $\sigma$                      &     149.7603   &    77.6237   &    71.0606  &     12.5094   &    70.5673    \\
Red $\chi^2$                        &     0.2127     &   0.0571     &   0.0479    &    0.0015     &   0.0472       \\
   \\
           Term ($n$)              &          Fe5270 &      Fe5335  &      Fe5406 &         Na $D$   \\
    $C_{\rm 0}$                    &        3.6687   &     4.2195   &      2.2520   &    1.3092    \\
   $\log\theta$                    &        26.8280  &     26.0111  &      23.4383  &    21.0015     \\
    $\sigma$                        &      (1.1715)  &    (0.8517)  &     (0.6402)   &  (1.2074)     \\
 $\log^2 \theta$                    &       44.6660  &     32.4921  &      67.7726  &   152.7714     \\
    $\sigma$                        &     (8.6353)   &   (6.2780)   &     (4.7191)   &  (8.8997)     \\
 $\log^3 \theta$                    &     -205.3421  &   -169.5044  &      -80.2579 &    412.8384     \\
    $\sigma$                        &     (81.3620)   &  (59.1509)   &     (44.4633)  &  (83.8526)      \\
$[{{\rm M}\over{\rm H}}]$          &      1.8747    &    2.2036    &      1.3076   &    1.1254     \\
    $\sigma$                        &     (0.1118)    &  (0.0813)    &     (0.0611)   &  (0.1152)     \\
$[{{\rm M}\over{\rm H}}]^2$        &       0.3043   &     0.3826   &      0.2548   &    0.2983     \\
    $\sigma$                        &     (0.0451)    &  (0.0328)    &     (0.0246)   &  (0.0465)     \\
$[{{\rm M}\over{\rm H}}]^3$        &      0.0158    &    0.0222    &      0.0164   &    0.0235      \\
    $\sigma$                        &     (0.0050)    &  (0.0036)    &     (0.0027)   &  (0.0051)     \\
$[{{\rm M}\over{\rm H}}]\log\theta$   &   6.5574    &    8.0558    &      8.4043   &   12.2936     \\
    $\sigma$                        &     (0.7041)    &  (0.5119)    &     (0.3848)   &  (0.7256)     \\
$[{{\rm M}\over{\rm H}}]^2\log\theta$ &   0.3685    &    0.6646    &       0.7870  &     1.4220     \\
    $\sigma$                        &     (0.1144)    &  (0.0831)    &     (0.0625)   &  (0.1179)     \\
$[{{\rm M}\over{\rm H}}]\log^2 \theta$ &  3.4790    &    2.5911    &      11.0017  &    33.9557     \\
    $\sigma$                        &      (2.5957)  &    (1.8871)  &     (1.4185)   &  (2.6752)      \\
Total $\sigma$                       &     81.8717   &    59.5215   &      44.7419  &    84.3780     \\
Red $\chi^2$                            &   0.0636   &     0.0336   &      0.0190   &    0.0675     \\
      \\
\hline\\
\enddata
\label{coeffsthetaSDSS}
\end{deluxetable}

\clearpage

\begin{deluxetable}{lrrrrrrr}
\tablecolumns{8}
\tablecaption{Same as Table \ref{coeffsthetaIDS}, but for independent variables $V-K$ and $[{{\rm M}\over {\rm H}}]$. }
\tablehead{
\colhead{Polynomial term} & \multicolumn{7}{c}{Lick index, $I$}                 \\ 
} 
\startdata
           Term ($n$)               &        Fe4383    &    Fe4531    &   Fe5015    &    Mg$_{\rm 1}$ &  Mg $b$     \\
    $C_{\rm 0}$                     &        2.0247    &   2.0280     &   0.0297    &    0.4264     &   1.9868    \\
   $V-K$                            &        -2.8895   &   -0.4509    &    4.5572   &    -0.6166    &   -3.5266    \\
    $\sigma$                        &        (1.3722)    &  (0.8606)     &  (0.8227)    &  (0.1854)    &  (0.4996)     \\
 $V-K^2$                            &       3.0594     &  1.1931      & -0.3775     &   0.2549      &  1.5847     \\
    $\sigma$                        &       (0.5385)     &  (0.3377)      & (0.3229)     & (0.0728)     & (0.1961)      \\
 $V-K^3$                            &        -0.4494   &   -0.1639    &    0.0199   &    -0.0236    &   -0.1178    \\
    $\sigma$                        &       (0.0657)     &  (0.0412)      & (0.0394)     & (0.0089)     & (0.0239)    \\
$[{{\rm M}\over{\rm H}}]$           &      -0.4489    &   0.9816     &   2.2984    &   -0.0033     &  -0.6624    \\
    $\sigma$                        &       (0.3320)     &  (0.2082)      &  (0.1991)     &  (0.0449)   &  (0.1209)    \\
$[{{\rm M}\over{\rm H}}]^2$         &      -0.1629    &   0.2169     &   0.7254    &   -0.0077     &  -0.1158      \\
    $\sigma$                        &        (0.0792)    &  (0.0497)     &  (0.0475)    &  (0.0107)     &  (0.0288)    \\
$[{{\rm M}\over{\rm H}}]^3$         &      -0.0373    &   0.0022     &   0.0540    &   -0.0012     &   0.0042     \\
    $\sigma$                        &       (0.0077)     &  (0.0048)      & (0.0046)     & (0.0010)      &  (0.0028)    \\
$[{{\rm M}\over{\rm H}}]V-K$        &      0.9985     &  0.6180      &  1.2306     &  -0.0144      &  0.1333    \\
    $\sigma$                        &       (0.1784)     &  (0.1119)      & (0.1069)     & (0.0241)      &  (0.0649)    \\
$[{{\rm M}\over{\rm H}}]^2V-K$      &      0.0059     &  0.0097      &  0.0696     &   0.0009      &  0.0749      \\
    $\sigma$                        &       (0.0158)     &  (0.0099)      & (0.0095)     & (0.0021)      &  (0.0057)     \\
$[{{\rm M}\over{\rm H}}]V-K^2$      &      -0.0728    &  -0.0308     &  -0.0576    &    0.0109     &   0.1277      \\
    $\sigma$                        &       (0.0292)     &  (0.0183)      & (0.0175)     & (0.0040)      &  (0.0106)     \\
Total $\sigma$                      &       1.5253     &  0.9566      &  0.9146     &   0.2061      &  0.5554    \\
Red $\chi^2$                        &       0.0149     &  0.0059      &  0.0054     &   0.0003      &  0.0020    \\
   \\
           Term ($n$)              &       Fe5270    &    Fe5335   &   Fe5406  &        Na $D$    \\ 
    $C_{\rm 0}$                    &       2.3521    &    2.1055   &   1.7732  &      2.0387    \\ 
   $V-K$                           &       -2.7555   &    -2.0371  &  -2.7317  &     -2.8668    \\ 
    $\sigma$                        &       (0.8051)   &   (0.5521)  &  (0.5025)   &  (0.9161)    \\ 
 $V-K^2$                           &       2.0821    &    1.7716   &  1.7435   &     1.2867    \\ 
    $\sigma$                        &      (0.3159)    &  (0.2166)   &  (0.1972)   &  (0.3595)    \\ 
 $V-K^3$                           &       -0.2818   &    -0.2483  &  -0.2162  &     -0.0768    \\ 
    $\sigma$                        &      (0.0386)    &  (0.0264)   &  (0.0241)   &  (0.0439)     \\ 
$[{{\rm M}\over{\rm H}}]$          &      0.4714    &    0.3801   &  -0.0986  &     -0.1966    \\ 
    $\sigma$                        &      (0.1948)    &  (0.1336)   &  (0.1216)   & (0.2217)    \\ 
$[{{\rm M}\over{\rm H}}]^2$        &      0.2452    &    0.2306   &   0.0954  &      0.0009    \\ 
    $\sigma$                        &      (0.0465)    &  (0.0319)   &   (0.0290)  &  (0.0529)    \\ 
$[{{\rm M}\over{\rm H}}]^3$        &      0.0158    &    0.0189   &   0.0119  &      0.0132    \\ 
    $\sigma$                        &      (0.0045)    &  (0.0031)   &   (0.0028)  &  (0.0051)    \\ 
$[{{\rm M}\over{\rm H}}]V-K$       &      0.6728    &    0.7581   &   0.5120  &      0.1566    \\ 
    $\sigma$                        &      (0.1046)    &  (0.0718)   &   (0.0653)  &  (0.1191)     \\ 
$[{{\rm M}\over{\rm H}}]^2V-K$     &      0.0212    &    0.0406   &   0.0440  &      0.0805    \\ 
    $\sigma$                        &      (0.0092)    &  (0.0063)   &   (0.0058)  &  (0.0105)    \\ 
$[{{\rm M}\over{\rm H}}]V-K^2$     &      -0.0390   &    -0.0413  &   0.0054  &      0.1307    \\ 
    $\sigma$                        &      (0.0172)    &  (0.0118)   &   (0.0107)  &  (0.0195)    \\ 
Total $\sigma$                      &      0.8949    &    0.6137   &   0.5586  &      1.0183     \\ 
Red $\chi^2$                        &      0.0051    &    0.0024   &   0.0020  &      0.0067     \\ 
      \\
\hline\\
\enddata
\label{coeffsvmkIDS}
\end{deluxetable}

\clearpage

\begin{deluxetable}{lrrrrrrrrrrr}
\tablecolumns{9}
\tablecaption{LTE Partial $I$ derivatives with respect to $T_{\rm eff}$ and $[{{\rm M}\over {\rm H}}]$, at each grid value of 
$T_{\rm eff}$ and $[{{\rm M}\over {\rm H}}]$, in index units and
in units of $\sigma$ for SDSS spectral resolution for our nine Lick-XMP indices.  
Sample entries only - the entire table is available electronically. }
\tablehead{
\colhead{} & \multicolumn{7}{c}{100 K $\times{{\partial I}\over{\partial T_{\rm eff}}}|_{[{{\rm M}/{\rm H}}]}$ }                 \\ 
\colhead{Lick index, $I$} & \multicolumn{7}{c}{$T_{\rm eff}$}                 \\ 
} 
\startdata
 Index & $[{{\rm M}\over {\rm H}}]$ &  6500 &  6000 &  5500 &  5250 &  5000 &  4750  &  4500 &  4250 \\
    Fe4383  &    -0.0 &                -0.1508 & -0.2423 & -0.3303 & -0.3706 & -0.4064 & -0.4355 & -0.4551 & -0.4610  \\
    Fe4383  &    -0.5 &                -0.1241 & -0.2129 & -0.2975 & -0.3358 & -0.3694 & -0.3961 & -0.4129 & -0.4157  \\
    Fe4383  &    -1.0 &                -0.0981 & -0.1841 & -0.2654 & -0.3017 & -0.3332 & -0.3575 & -0.3717 & -0.3714  \\
    Fe4383  &    -1.5 &                -0.0727 & -0.1560 & -0.2340 & -0.2685 & -0.2978 & -0.3198 & -0.3313 & -0.3281   \\
    Fe4383  &    -2.0 &                -0.0480 & -0.1286 & -0.2034 & -0.2360 & -0.2633 & -0.2830 & -0.2919 & -0.2857  \\
\\
\hline
\\
\colhead{Lick index, $I$} & \multicolumn{7}{c}{100 K $\times{{\partial I}\over{\partial T_{\rm eff}}}|_{[{{\rm M}/{\rm H}}]}/\sigma$  }                 \\ 
\\
\hline
\\
Index & $[{{\rm M}\over {\rm H}}]$ &  6500 &   6000 &  5500 &  5250 &  5000 &  4750  &  4500 &  4250 \\
    Fe4383  &    -0.0 &               -0.0010 & -0.0016 & -0.0022 & -0.0025 & -0.0027 & -0.0029 & -0.0030 & -0.0031   \\
    Fe4383  &    -0.5 &               -0.0008 & -0.0014 & -0.0020 & -0.0022 & -0.0025 & -0.0026 & -0.0028 & -0.0028   \\
    Fe4383  &    -1.0 &               -0.0007 & -0.0012 & -0.0018 & -0.0020 & -0.0022 & -0.0024 & -0.0025 & -0.0025  \\
    Fe4383  &    -1.5 &               -0.0005 & -0.0010 & -0.0016 & -0.0018 & -0.0020 & -0.0021 & -0.0022 & -0.0022    \\
    Fe4383  &    -2.0 &               -0.0003 & -0.0009 & -0.0014 & -0.0016 & -0.0018 & -0.0019 & -0.0019 & -0.0019     \\
 
\\
\hline
\\
\colhead{Lick index, $I$} & \multicolumn{7}{c}{$0.5\times {{\partial I}\over{\partial [{{\rm M}\over {\rm H}}]}}|_{T_{\rm eff}}$ }                 \\ 
\\
\hline
\\
 Index  & $[{{\rm M}\over {\rm H}}]$ &  6500 &   6000 &  5500 &  5250 &  5000 &  4750  &  4500 &  4250 \\
    Fe4383  &    -0.0 &                 0.3247 &  0.4664 &  0.6237 &  0.7092  & 0.7999 &  0.8964 &  0.9994  & 1.1098  \\
    Fe4383  &    -0.5 &                 0.3942 &  0.5326 &  0.6863 &  0.7699  & 0.8586 &  0.9530 &  1.0539  & 1.1619   \\
    Fe4383  &    -1.0 &                 0.4342 &  0.5693 &  0.7195 &  0.8012  & 0.8879 &  0.9803 &  1.0789  & 1.1846  \\
    Fe4383  &    -1.5 &                 0.4449 &  0.5767 &  0.7233 &  0.8031  & 0.8879 &  0.9781 &  1.0745  & 1.1779  \\
    Fe4383  &    -2.0 &                 0.4261 &  0.5547 &  0.6977 &  0.7756  & 0.8584 &  0.9465 &  1.0408  & 1.1418  \\
\\ 
\hline
\\
\colhead{Lick index, $I$} & \multicolumn{7}{c}{$0.5\times {{\partial I}\over{\partial [{{\rm M}\over {\rm H}}]}}|_{T_{\rm eff}}/\sigma$}                 \\ 
\\
\hline
\\
 Index  & $[{{\rm M}\over {\rm H}}]$ &  6500 &   6000 &  5500 &  5250 &  5000 &  4750  &  4500 &  4250 \\
    Fe4383  &    -0.0 &                 0.0022 &  0.0031 &  0.0042  & 0.0047 &  0.0053  & 0.0060 &  0.0067  & 0.0074   \\
    Fe4383  &    -0.5 &                 0.0026 &  0.0036 &  0.0046  & 0.0051 &  0.0057  & 0.0064 &  0.0070  & 0.0078   \\
    Fe4383  &    -1.0 &                 0.0029 &  0.0038 &  0.0048  & 0.0053 &  0.0059  & 0.0065 &  0.0072  & 0.0079   \\
    Fe4383  &    -1.5 &                 0.0030 &  0.0039 &  0.0048  & 0.0054 &  0.0059  & 0.0065 &  0.0072  & 0.0079   \\
    Fe4383  &    -2.0 &                 0.0028 &  0.0037 &  0.0047  & 0.0052 &  0.0057  & 0.0063 &  0.0069  & 0.0076   \\
\hline\\
\enddata
\label{partialteffs}
\end{deluxetable}

\clearpage

\begin{deluxetable}{lrrrrrrrrrrr}
\tablecolumns{9}
\tablecaption{Same as Table \ref{partialteffs}, but with respect to $V-K$ and $[{{\rm M}\over {\rm H}}]$, at each grid value of 
$V-K$ for $[{{\rm M}\over {\rm H}}]=0.0$ models and $[{{\rm M}\over {\rm H}}]$.  Note that $V-K$ is
$[{{\rm M}\over {\rm H}}]$-dependent and the $V-K$ values used to label the columns are only valid for 
$[{{\rm M}\over {\rm H}}]=0.0$ (Fig. \ref{vmkteff} gives an indication of the $[{{\rm M}\over {\rm H}}]$-dependence
on $V-K$.)  Sample entries only - the entire table is available electronically.   }
\tablehead{
\colhead{} & \multicolumn{7}{c}{0.25 mag $\times {{\partial I}\over{\partial (V-K)}}|_{[{{\rm M}/{\rm H}}]}$}                 \\ 
\colhead{Lick index, $I$} & \multicolumn{7}{c}{$(V-K)$ at $[{{\rm M}/{\rm H}}]=0.0$}                 \\ 
} 
\startdata
 Index & $[{{\rm M}\over {\rm H}}]$ &  0.887   &   1.211   &  1.578   &  1.792   &  2.017   &  2.290    &  2.585   \\ 
    Fe4383  &    -0.0 &                0.6126  & 0.8463  & 1.0189 &  1.0744 &  1.0969 &  1.0748 &  0.9900  \\ 
    Fe4383  &    -0.5 &                0.5123  & 0.7459  & 0.9227 &  0.9801 &  1.0083 &  0.9960 &  0.9239  \\ 
    Fe4383  &    -1.0 &                0.4144  & 0.6498  & 0.8299 &  0.8884 &  0.9203 &  0.9136 &  0.8495  \\ 
    Fe4383  &    -1.5 &                0.3137  & 0.5543  & 0.7361 &  0.7977 &  0.8326 &  0.8309 &  0.7770  \\ 
    Fe4383  &    -2.0 &                0.2136  & 0.4555  & 0.6419 &  0.7063 &  0.7449 &  0.7487 &  0.7029  \\ 
\\
\hline
\\
\colhead{Lick index, $I$} & \multicolumn{7}{c}{0.25 mag $\times {{\partial I}\over{\partial (V-K)}}|_{[{{\rm M}\over {\rm H}}]}/\sigma$}                 \\ 
\\
\hline
\\
 Index & $[{{\rm M}\over {\rm H}}]$ &  0.887   &   1.211   &  1.578   &  1.792   &  2.017   &  2.290    &  2.585   \\ 
    Fe4383  &    -0.0 &               -2.8131 & -2.2252 & -1.6230 & -1.3032 & -0.9918 & -0.6481 & -0.3189 \\ 
    Fe4383  &    -0.5 &               -2.9024 & -2.3225 & -1.7227 & -1.4147 & -1.1006 & -0.7738 & -0.4398 \\ 
    Fe4383  &    -1.0 &               -2.9874 & -2.4089 & -1.8095 & -1.5086 & -1.1932 & -0.8744 & -0.5388 \\ 
    Fe4383  &    -1.5 &               -3.0797 & -2.4949 & -1.9022 & -1.5992 & -1.2884 & -0.9726 & -0.6457 \\ 
    Fe4383  &    -2.0 &               -3.1715 & -2.5911 & -1.9978 & -1.6955 & -1.3860 & -1.0757 & -0.7484 \\ 
\\
\hline
\\
\colhead{Lick index, $I$} & \multicolumn{7}{c}{$0.5\times {{\partial I}\over{\partial [{{\rm M}\over {\rm H}}]}}|_{(V-K)}$}                 \\ 
\\
\hline
\\
 Index & $[{{\rm M}\over {\rm H}}]$ &  0.887   &   1.211   &  1.578   &  1.792   &  2.017   &  2.290    &  2.585   \\ 
    Fe4383  &    -0.0 &                -1.0540 & -0.8495 & -0.6315 & -0.5151 & -0.3944 & -0.2651 & -0.1258  \\ 
    Fe4383  &    -0.5 &                -0.9953 & -0.7911 & -0.5733 & -0.4570 & -0.3365 & -0.2074 & -0.0683  \\ 
    Fe4383  &    -1.0 &                -0.9658 & -0.7618 & -0.5442 & -0.4281 & -0.3077 & -0.1787 & -0.0398  \\ 
    Fe4383  &    -1.5 &                -0.9654 & -0.7616 & -0.5443 & -0.4283 & -0.3080 & -0.1792 & -0.0405  \\ 
    Fe4383  &    -2.0 &                -0.9940 & -0.7905 & -0.5734 & -0.4575 & -0.3374 & -0.2088 & -0.0702  \\ 
\\ 
\hline
\\
\colhead{Lick index, $I$} & \multicolumn{7}{c}{$0.5\times {{\partial I}\over{\partial [{{\rm M}\over {\rm H}}]}}|_{(V-K)}/\sigma$}                 \\ 
\\
\hline
\\
 Index & $[{{\rm M}\over {\rm H}}]$ &  0.887   &   1.211   &  1.578   &  1.792   &  2.017   &  2.290    &  2.585   \\ 
    Fe4383  &    -0.0 &                -0.7299 & -0.5883 & -0.4373 & -0.3567 & -0.2731 & -0.1836 & -0.0871 \\ 
    Fe4383  &    -0.5 &                -0.6893 & -0.5479 & -0.3970 & -0.3165 & -0.2330 & -0.1436 & -0.0473 \\ 
    Fe4383  &    -1.0 &                -0.6689 & -0.5276 & -0.3769 & -0.2965 & -0.2131 & -0.1238 & -0.0276 \\ 
    Fe4383  &    -1.5 &                -0.6686 & -0.5274 & -0.3769 & -0.2966 & -0.2133 & -0.1241 & -0.0280 \\ 
    Fe4383  &    -2.0 &                -0.6884 & -0.5474 & -0.3971 & -0.3169 & -0.2337 & -0.1446 & -0.0486 \\ 
\hline\\
\enddata
\label{partialvmks}
\end{deluxetable} 

\end{document}